\documentclass[preprint,showpacs,preprintnumbers,eqsecnum,floatfix]{revtex4}
\usepackage{graphicx}% Include figure files
\usepackage{dcolumn}% Align table columns on decimal point
\usepackage{bm}% bold math
\def\lesssim{\ \raise.3ex\hbox{$<$}\kern-0.8em\lower.7ex\hbox{$\sim$}\ }
\def\gesim{\ \raise.3ex\hbox{$>$}\kern-0.8em\lower.7ex\hbox{$\sim$}\ }
\font\scripti=cmmi7
\font\scriptscripti=cmmi5
\def\sib#1{\setbox0 = \hbox{\scripti #1}
  \kern-.02em\copy0\kern-\wd0
  \kern.04em\box0} % script italic bold 
\def\ssib#1{\setbox0 = \hbox{\scriptscripti #1}
  \kern-.02em\copy0\kern-\wd0
  \kern.04em\box0} % scriptscript italic bold
\font\tenib=cmmib10 % italic bold for math
\skewchar\tenib='177 \skewchar\tenib='177 \skewchar\tenib='177
\def\pbold#1{\setbox0 = \hbox{$ #1 $}
  \kern-.022em\copy0\kern-\wd0
  \kern.011em\copy0\kern-\wd0
  \kern.011em\copy0\kern-\wd0
  \kern.011em\copy0\kern-\wd0
  \kern.011em\box0} % poorman's bold
\begin{document}
%\preprint{APS/123-QED}
\title{Superfluid density and condensate fraction in the BCS-BEC crossover regime at finite temperatures}
\author{N. Fukushima$^{1}$, Y. Ohashi$^{1,2}$, E. Taylor$^{3}$, and A. Griffin$^{3}$}
\affiliation{
$^1$Institute of Physics, University of Tsukuba, Tsukuba, Ibaraki 305, Japan, \\
$^2$Faculty of Science and Technology, Keio University, Hiyoshi, Yokohama, 223, Japan,
\\
$^3$ Department of Physics, University of Toronto, Toronto, Ontario, Canada M5S 1A7}
\date{\today}
\begin{abstract}
The superfluid density is a fundamental quantity describing the response to a rotation as well as in two-fluid collisional hydrodynamics. We present extensive calculations of the superfluid density $\rho_s$ in the BCS-BEC crossover regime of a uniform superfluid Fermi gas at finite temperatures.We include strong-coupling or fluctuation effects on these quantities within a Gaussian approximation. We also incorporate the same fluctuation effects into the BCS single-particle excitations described by the superfluid order parameter $\Delta$ and Fermi chemical potential $\mu$, using the Nozi\`eres and Schmitt-Rink (NSR) approximation. This treatment is shown to be necessary for consistent treatment of $\rho_s$ over the entire BCS-BEC crossover. We also calculate the condensate fraction $N_c$ as a function of the temperature, a quantity which is quite different from the superfluid density $\rho_s$. We show that the mean-field expression for the condensate fraction $N_c$ is a good approximation even in the strong-coupling BEC regime. Our numerical results show how $\rho_s$ and $N_c$ depend on temperature, from the weak-coupling BCS region to the BEC region of tightly-bound Cooper pair molecules. In a companion paper by the authors (cond-mat/0609187), we derive an equivalent expression for $\rho_s$ from the thermodynamic potential, which exhibits the role of the pairing fluctuations in a more explicit manner.
\end{abstract}
\maketitle
%
%%%%%%%%%%%%%%%%%%%%%%%%%%%%%%%%%%%%%%%%%%%%%%%%%%%%%%%%%%%%%%%%%%%%%%
\section{Introduction}
In the last few years, the BCS-BEC crossover in two-component Fermi superfluids has become a central topic in ultracold atom physics\cite{Levin,Nozieres,Randeria,Ohashi2}. This crossover is of especial interest since the superfluidity continuously changes from the weak-coupling BCS-type to the Bose-Einstein condensation (BEC) of tightly bound Cooper pairs, as one increases the strength of a pairing interaction\cite{Jin}. Thus the BCS-BEC crossover enables us to study fermion superfluidity and boson superfluidity in a unified manner.
\par
The superfluid density $\rho_s$ is a fundamental quantity which describes the response of a superfluid which arises from a BEC\cite{Pines}. The superfluid density was first introduced by Landau as part of the two-fluid theory of superfluid $^4$He\cite{Landau}. At $T=0$, the value of $\rho_s$ always equals the total carrier density $n$. (In the BCS-BEC crossover, $n$ is the total number density of fermions.) This property is satisfied in both the Fermi superfluids and Bose superfluids, irrespective of the strength of the interaction between particles. This is quite different from what is called the condensate fraction $N_c$, which describes the number of the Bose-condensed particles\cite{Landau,Yang}. For example, in superfluid $^4$He, only about 10\% of atoms are Bose-condensed even at $T=0$, due to the strong repulsion between the $^4$He atoms [For a review, see Ref. \cite{Griffin}.]. In contrast, all the atoms contribute to the superfluid density at $T=0$, namely $\rho_s(T=0)=n$. 
\par
In a companion paper, we have discussed some analytical results for the superfluid density $\rho_s$ in the BCS-BEC crossover regime of a uniform superfluid Fermi gas\cite{Ed}. Going past the weak-coupling BCS theory, we derived an expression for $\rho_s$ in the Gaussian fluctuation level in terms of the fluctuations in the Cooper channel. The resulting expression for the normal fluid density $\rho_n\equiv n-\rho_s$ is given by the sum of the usual BCS normal fluid density $\rho_n^F$ and a bosonic fluctuation contribution $\rho_n^B$. While the superfluid density from fermions dominates in the weak-coupling BCS regime, the bosonic fluctuation contribution $\rho_n^B$ becomes dominant in the strong-coupling BEC regime. Since $\rho_n^B$ is absent in the mean-field BCS theory, inclusion of fluctuations in the Cooper channel is clearly essential in considering the superfluid density in the BCS-BEC crossover.
\par
In Ref. \cite{Ed}, our expression for $\rho_s$ was obtained using the thermodynamic potential in the presence of a superfluid flow. In the present paper, we derive a second expression for $\rho_s$ by calculating the effect of pairing fluctuations on the single-particle Green's function with a supercurrent. In this paper, we use this expression to numerically calculate $\rho_s$ in the entire BCS-BEC crossover regime at finite temperatures. However, it can be shown that this expression for $\rho_s$ (given in Sec. III) is equivalent to the result derived in Ref. \cite{Ed}. In calculating the superfluid order parameter $\Delta$ as well as Fermi chemical potential $\mu$, we include the effect of the pairing fluctuations following the approach given in Ref. \cite{Ohashi2}. This self-consistent treatment of $\Delta$ and $\mu$ is crucial in calculating $\rho_s$ as a function of the temperature in the BCS-BEC crossover.
\par
Besides the superfluid density, we also calculate the condensate fraction $N_c$ describing the number of Bose-condensed particles\cite{Yang}. $N_c$ is of special interest in superfluid Fermi gases, since it can be observed experimentally. Indeed, a finite value of $N_c$ is the signature of the BCS-BEC superfluid phase\cite{Jin}. In this paper, we show that strong-coupling pair fluctuations have little effect on $N_c$ in the BCS-BEC crossover. We note that $N_c$ has been recently calculated at $T=0$ within a simple mean-field BCS approach\cite{Luca}, and using Monte Carlo techniques\cite{Gio}. In this paper, we present detailed results for $N_c$ at finite temperatures in the BCS-BEC crossover, based on the NSR theory of fluctuations\cite{Nozieres}. 
\par
The present paper is organized as follows. In Sec. II, the BCS Green's functions are solved numerically for the superfluid order parameter $\Delta$ and chemical potential $\mu$ self-consistently. This is done for the entire BCS-BEC crossover region and at finite temperatures. In Sec. III, we calculate the superfluid density $\rho_s$ treating the strong-coupling pair fluctuation effects within a Gaussian approximation\cite{Nozieres,Randeria,Ohashi2,Engelbrecht}. Numerical results for $\rho_s$ are presented in Sec. IV. In Sec. V, we define and calculate the condensate fraction $N_c$. 
\par
Throughout this paper, we take $\hbar=k_B=1$. We also set the volume $V=1$, so that the number of atoms $N$ and the number density $n$ are the same.
\par
\section{BCS-BEC crossover in the superfluid phase}
\par
In superfluid Fermi gases, current experiments make use of a broad Feshbach resonance to tune the magnitude and sign of the pairing interaction\cite{Jin}. In this case, the superfluid properties can be studied by using the single-channel BCS model, described by the Hamiltonian,
\begin{equation}
H=\sum_{{\bf p},\sigma}\xi_{\bf p}
c_{{\bf p}\sigma}^\dagger c_{{\bf p}\sigma}
-U\sum_{{\bf p},{\bf p}',{\bf q}}
c_{{\bf p}+{\bf q}\uparrow}^\dagger c_{{\bf p}'-{\bf q}\downarrow}^\dagger
c_{{\bf p}'\downarrow} c_{{\bf p}\uparrow}.
\label{eq.1}
\end{equation}
Here, $c_{{\bf p}\sigma}^\dagger$ is a creation operator of a Fermi atom with pseudo-spin $\sigma=\uparrow,\downarrow$ (which describe the two atomic hyperfine states.). The Fermi atoms have kinetic energy $\xi_{\bf p}=\varepsilon_{\bf p}-\mu=p^2/2m-\mu$, measured from the Fermi chemical potential $\mu$. $-U$ describes a pairing interaction between different Fermi atoms. The magnitude and sign of $U$ can be tuned using the Feshbach resonance using an external magnetic field. The weak-coupling BCS limit corresponds to $U\to +0$. We only consider a uniform gas in this paper. 
\par
Nozi\`eres and Schmitt-Rink (NSR) first discussed the BCS-BEC crossover behavior of the Hamiltonian in Eq. (\ref{eq.1}) to determine $T_{\rm c}$\cite{Nozieres}, which we briefly summarize (see also Ref. \cite{Ohashi1}), based on a Gaussian approximation for pair fluctuations\cite{Randeria}. The NSR theory gives two coupled equations, one for $T_{\rm c}$, and one giving $\mu$ as a function of $N$. The equation for $T_{\rm c}$, given by the Thouless criterion, has the same form as the weak-coupling BCS gap equation at $T_{\rm c}$,
\begin{equation}
1=U\sum_{\bf p}{1 \over 2\xi_{\bf p}}\tanh {\beta \over 2}\xi_{\bf p},
\label{eq.5}
\end{equation} 
where $\beta=1/T$. In the weak-coupling BCS theory, one finds that the equation of state gives $\mu=\varepsilon_{\rm F}$, where $\varepsilon_{\rm F}$ is the Fermi energy. However, this result for $\mu$ is not valid in the BCS-BEC crossover regime where we must include strong-coupling effects due to pairing fluctuations. In the NSR theory, this deviation is determined by solving the equation for the number of fermions\cite{Nozieres,Ohashi2,Ohashi1,notez}, within a $t$-matrix approximation in the particle-particle channel,
\begin{equation}
N=2\sum_{\bf p}f(\xi_{\bf p})-{1 \over \beta}
{\partial \over \partial \mu}\sum_{{\bf q},i\nu}
\ln [1-U\Pi({\bf q},i\nu_n)].
\label{eq.6}
\end{equation}
Here $f(\xi_{\bf p})$ is the Fermi distribution function, and $\Pi({\bf q},i\nu_n)$ describes non-interacting fluctuations in the Cooper channel,
\begin{equation}
\Pi({\bf q},i\nu_n)=
\sum_{\bf p}
{1-f(\xi_{{\bf p}+{\bf q}/2})-f(\xi_{{\bf p}-{\bf q}/2})
\over 
\xi_{{\bf p}+{\bf q}/2}+\xi_{{\bf p}+{\bf q}/2}-i\nu_n
}.
\label{eq.7}
\end{equation}
Here, $\nu_n$ is the boson Matsubara frequency describing bosonic fluctuations. The second term in Eq. (\ref{eq.6}) gives the number of Cooper pairs which are formed, and which become stable as we go through the Feshbach resonance. In the extreme BEC limit, all the fermions form such stable molecules.
\par
The NSR theory to the superfluid phase has been studied below $T_{\rm c}$\cite{Ohashi2,Engelbrecht}. We briefly review this approach. It is convenient to introduce the two-component Nambu field operator\cite{Schrieffer,Maki},
\begin{eqnarray}
\Psi_{\bf p}=
\left(
\begin{array}{c}
c_{{\bf p}\uparrow}\\
c_{-{\bf p}\downarrow}^\dagger
\end{array}
\right).
\label{eq.2}
\end{eqnarray}
In this standard Nambu representation, Eq. (\ref{eq.1}) can be rewritten as\cite{Takada},
\begin{eqnarray}
H={\Delta^2 \over U}+\sum_{\bf p}\xi_{\bf p}+\sum_{\bf p}\Psi_{\bf p}^\dagger[\xi _{\bf p}\tau_3-\Delta\tau_1]\Psi_{\bf p}
-{U \over 4}\sum_{\bf q}[\rho_{1,{\bf q}}\rho_{1,-{\bf q}}+\rho_{2,{\bf q}}\rho_{2,-{\bf q}}].
\label{eq.3}
\end{eqnarray}
The mean-field BCS approximation is described by the first three terms of Eq. (\ref{eq.3}), with the $2\times 2$-matrix Nambu-Gor'kov propagator\cite{Schrieffer}
\begin{eqnarray}
{\hat G}_0({\bf p},i\omega_m)
&=&
{1 \over i\omega_m-\xi_{\bf p}\tau_3+\Delta\tau_1}
\nonumber
\\
&=&
{i\omega_m+\xi_{\bf p}\tau_3-\Delta\tau_1 \over (i\omega_m)^2-E_{\bf p}^2},\label{eq.9}
\end{eqnarray}
where $\omega_m$ is the fermion Matsubara frequency associated with the fermions and $\tau_j$ are the Pauli operators. $E_{\bf p}=\sqrt{\xi_{\bf p}^2+\Delta^2}$ describes Bogoliubov single-particle excitations, associated with the breaking of a Cooper pair. 
The order parameter $\Delta\equiv U\sum_{\bf p}\langle c_{{\bf p}\uparrow}^\dagger c_{-{\bf p}\downarrow}^\dagger\rangle$ (taken to be real) corresponds to an off-diagonal mean-field and is related to the thermal Green's function ${\hat G}_0({\bf p},i\omega_m)$ as
\begin{equation}
\Delta={U \over \beta}\sum_{{\bf p},\omega_m}
G_0^{12}({\bf p},i\omega_m).
\label{eq.8}
\end{equation}
Substituting Eq. (\ref{eq.9}) into Eq. (\ref{eq.8}), we obtain the well-known gap equation,
\begin{equation}
\Delta=U\sum_{\bf p}{\Delta \over 2E_{\bf p}}\tanh{\beta E_{\bf p} \over 2}.
\label{eq.10}
\end{equation}
Equation (\ref{eq.10}) is an implicit equation determining $\Delta$ as a function of $\mu$ and $T$. In the path integral formalism, the result in Eq. (\ref{eq.10}) emerges as the saddle point solution of Eq. (\ref{eq.3})\cite{Randeria}. Equation (\ref{eq.10}) reduces to Eq. (\ref{eq.5}) when $\Delta=0$, which defines $T_{\rm c}$.
\par
$\rho_{j,{\bf q}}$ ($j=1,2$) represent the generalized density operators, defined by\cite{note1}
\begin{eqnarray}
\rho_{j,{\bf q}}=\sum_{{\bf p}}{\Psi^{\dagger}_{{\bf p+q}/2}\tau_{j}\Psi_{{\bf p-q}/2}}.
\label{eq.4}
\end{eqnarray}
Writing out the operators, one sees that $\rho_{1,{\bf q}}$ and $\rho_{2,{\bf q}}$ describe amplitude and phase fluctuations of the Cooper-pair order parameter $\Delta$, respectively. As shown in Eq. (\ref{eq.3}), the interaction separates into amplitude fluctuations ($\rho_{1,{\bf q}}\rho_{1,-{\bf q}}$) and phase fluctuations ($\rho_{2,{\bf q}}\rho_{2,-{\bf q}}$). 
\par

%%%%%%%%%%%%%%%%%%%%%%%%%%%%%%%%%%%%

\begin{figure}
\includegraphics[width=15cm,height=3cm]{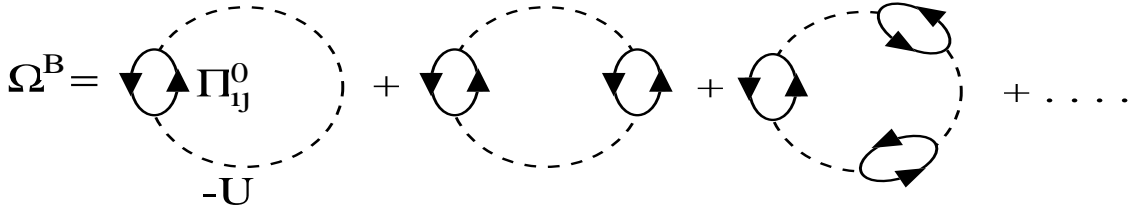}% 
\caption{Fluctuation contribution to the thermodynamic potential $\Omega$. The solid line describes the single particle matrix Green's function ${\hat G}_0$, and the dashed line represents the pairing interaction $-U$. $\Pi_{ij}^0$ ($i,j=1,2$) are the generalized correlation functions, given by Eqs. (\ref{eq.16})-(\ref{eq.18}).
\label{fig1}
}
\end{figure}

%%%%%%%%%%%%%%%%%%%%%%%%%%%%%%%%%%%%

\par
The chemical potential $\mu$ is determined by the equation for the total number of particles $N$. This equation is conveniently obtained from the thermodynamic identity $N=-\partial \Omega/\partial \mu$. We calculate the thermodynamic potential $\Omega$ perturbatively, taking into account the interaction in Eq. (\ref{eq.3}). In the Nambu representation, the fluctuation contribution to $\Omega$ is diagrammatically described by Fig. \ref{fig1}. As discussed in Refs. \cite{Ohashi2,Engelbrecht}, one finds [See Appendix A.]
\begin{eqnarray}
N=N_F^0-{1 \over 2\beta}{\partial \over \partial\mu}
\sum_{{\bf q},\nu_n}
\ln
{\rm det}
\Bigl[
1+U{\hat \Xi}({\bf q},i\nu_n)
\Bigr],
\label{eq.23}
\end{eqnarray}
where the first term
\begin{equation}
N_F^0=
\sum_{\bf p}
\Bigl[
1-{\xi_{\bf p} \over E_{\bf p}}\tanh{\beta \over 2}E_{\bf p}
\Bigr]
\label{eq.24}
\end{equation}
comes from the mean-field thermodynamic potential $\Omega^F$ describing BCS Fermi quasiparticles. The second term in Eq. (\ref{eq.23}) describes the thermodynamic potential contribution $\Omega^B$ from bosonic collective pair fluctuations. ${\hat \Xi}$ is a $2\times 2$-matrix correlation function defined in Eq. (\ref{eq.20}). We note that the $\mu$-derivative in Eq. (\ref{eq.23}) only acts on the chemical potential in the kinetic energy $\xi_{\bf p}$. Equation (\ref{eq.23}) reduces to the NSR result in Eq. (\ref{eq.6}) at $T_{\rm c}$ [where $\Delta=0$ and $\Xi_{11}({\bf q},i\nu_n)=\Xi_{22}({\bf q},-i\nu_n)=-\Pi({\bf q},i\nu_n)$]. See Appendix A for more details.
\par
We solve Eq. (\ref{eq.10}) together with Eq. (\ref{eq.23}) to give $\Delta$ and $\mu$ self-consistently. As usual, we need to introduce a high energy cutoff $\omega_c$ in these coupled equations. This cutoff can be formally eliminated by introducing the two-body $s$-wave scattering length $a_s$\cite{Randeria},
\begin{equation}
{4\pi a_s \over m}\equiv -{U \over 1-U\sum_{\bf p}^{\omega_c}{1 \over 2\varepsilon_{\bf p}}}.
\label{eq.25}
\end{equation}
Using $a_s$ in place of $U$, one can rewrite Eqs. (\ref{eq.10}) and (\ref{eq.23}) in the form
\begin{eqnarray}
1=-{4\pi a_s \over m}\sum_{\bf p}
\Bigl[
{1 \over 2E_{\bf p}}\tanh{\beta \over 2}E_{\bf p}-{1 \over 2\varepsilon_{\bf p}}
\Bigr],
\label{eq.26}
\end{eqnarray}
\begin{eqnarray}
N=N_{\rm F}^0
-{1 \over 2\beta}
{\partial \over \partial \mu}
\sum_{{\bf q},\nu_n}
\ln
{\rm det}
\Bigl[
1-{4\pi a_s \over m}
[{\hat \Xi}({\bf q},i\nu_n)+{1 \over 2\varepsilon_{\bf p}}]
\Bigr],
\label{eq.27}
\end{eqnarray}
where the momentum sums are now no longer divergent. The weak-coupling BCS regime and the strong-coupling BEC regime are, respectively, given by $(k_{\rm F}a_s)^{-1}\lesssim -1$ and $(k_{\rm F}a_s)^{-1}\gesim 1$ (where $k_{\rm F}$ is the Fermi momentum). The region $-1\lesssim(k_{\rm F}a_s)^{-1}\lesssim 1$ is referred to as the ``crossover regime." In Appendix B, we show that Eqs. (\ref{eq.26}) and (\ref{eq.27}) are the limiting results obtained from a two-channel coupled fermion-boson model\cite{Ohashi2} when the Feshbach resonance is broad. 

%%%%%%%%%%%%%%%%%%%%%%%%%%%%%%%%%%%%

\begin{figure}
\includegraphics[width=8cm,height=10cm]{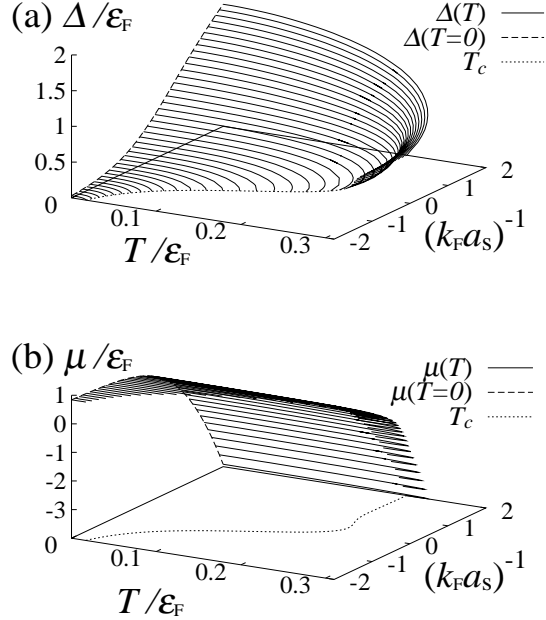}% 
\caption{
(a) Off-diagonal mean-field $\Delta$, and (b) Fermi chemical potential $\mu$ in the BCS-BEC crossover. The pairing interaction is measured in terms of the inverse of the two-body scattering length $a_s$, normalized by the Fermi momentum $k_{\rm F}$. In these panels, the dotted line shows $T_{\rm c}$ as a function of $(k_{\rm F}a_s)^{-1}$. In the strong-coupling regime, the apparent first order behavior of the phase transition is an artifact of the NSR Gaussian treatment of pairing fluctuations (see text).
\label{fig2}
}
\end{figure}

%%%%%%%%%%%%%%%%%%%%%%%%%%%%%%%%%%%%

\begin{figure}
\includegraphics[width=8cm,height=10cm]{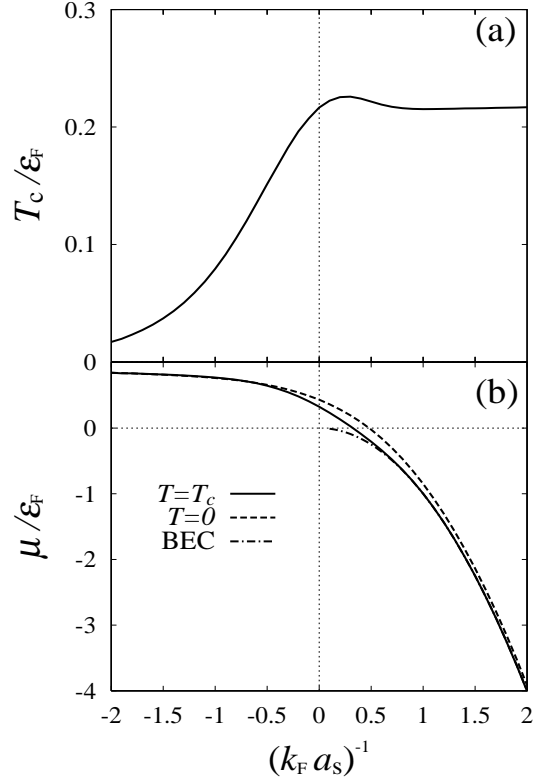}% 
\caption{
(a) Superfluid phase transition temperature $T_{\rm c}$, and (b) chemical potential $\mu(T=T_{\rm c})$ in the BCS-BEC crossover. In panel (b), $\mu$ at $T=0$ is also shown. The curve `BEC' gives the strong-coupling BEC limit, where one finds $\mu=-1/2ma_s^2$.
\label{fig3}
}
\end{figure}

%%%%%%%%%%%%%%%%%%%%%%%%%%%%%%%%%%%%

\begin{figure}
\includegraphics[width=10cm,height=6cm]{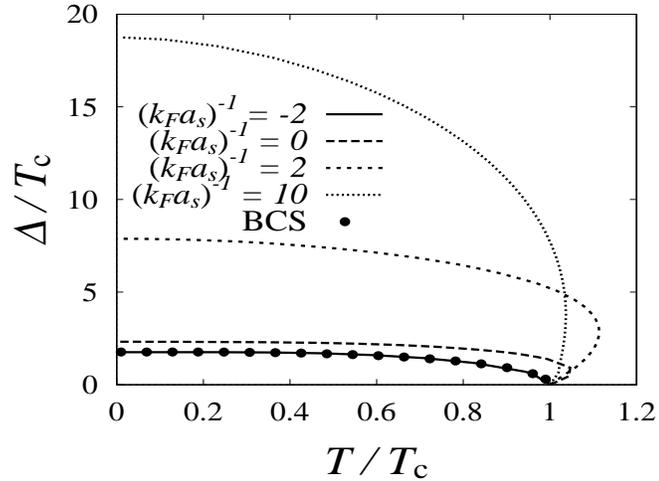}% 
\caption{
Calculated values of the superfluid order parameter $\Delta$ as a function of temperature. `BCS' labels the weak-coupling BCS limit. The bendover near $T_{\rm c}$ is an artifact of our NSR Gaussian treatment of fluctuations.
\label{fig4}
}
\end{figure}

%%%%%%%%%%%%%%%%%%%%%%%%%%%%%%%%%%%%

%%%%%%%%%%%%%%%%%%%%%%%%%%%%%%%%%%%%

\begin{figure}
\includegraphics[width=10cm,height=6cm]{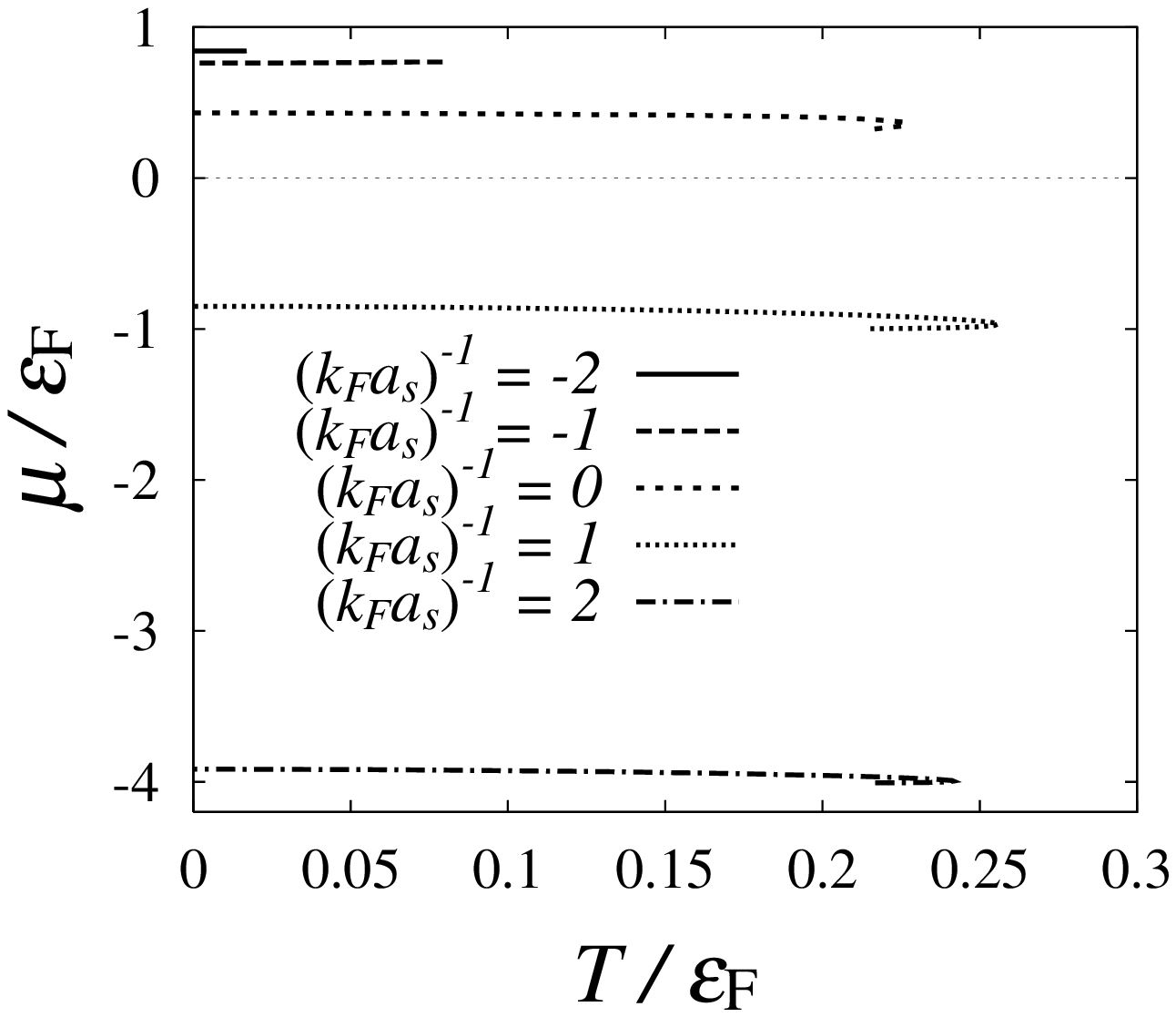}% 
\caption{
Fermi chemical potential $\mu$ as a function of temperature. For comparison, in the unitarity limit [$(k_{\rm F}a_s)^{-1}=0$], MC results\cite{Gio,ADD} give $\mu/\varepsilon_{\rm F}=0.44\sim 0.49$, and an improved version of NSR theory\cite{HuiHu} gives $\mu/\varepsilon_{\rm F}=0.4\sim 0.47$ just below $T_{\rm c}$.
\label{fig5}
}
\end{figure}

%%%%%%%%%%%%%%%%%%%%%%%%%%%%%%%%%%%%

\par
Figure \ref{fig2} shows our self-consistent solutions of the coupled equations (\ref{eq.26}) and (\ref{eq.27}) in the BCS-BEC crossover at finite temperatures. These reproduce the NSR results for $T_{\rm c}$\cite{Randeria}, as shown explicitly in Fig. \ref{fig3}. 
\par
When one enters the crossover region, Fig. \ref{fig4} shows that the order parameter $\Delta$ deviates from the weak-coupling BCS result. In the strong-coupling BEC regime, although the superfluid phase transition approaches the value $T_{\rm c}=0.218T_{\rm F}$\cite{Randeria}, $\Delta(T=0)$ continues to increase. As a result, the ratio $2\Delta(T=0)/T_{\rm c}$ in the BEC regime is larger than the weak-coupling BCS universal constant $2\Delta(T=0)/T_{\rm c}=3.54$. We recall that on the BEC side of the crossover, where $\mu$ is negative, the energy gap is not equal to $\Delta$\cite{Ohashi2,Engelbrecht}.
\par
The chemical potential is strongly affected by fluctuations in the Cooper channel, and becomes negative in the BEC regime, as shown in Figs. \ref{fig2}(b) and \ref{fig3}(b). In the strong-coupling BEC regime, $\mu$ approaches $\mu=-1/2ma_s^2$\cite{Leggett}. Although $\mu$ strongly depends on the magnitude of the interaction, Fig. \ref{fig5} shows that the temperature dependence of $\mu$ is very weak in the entire BCS-BEC crossover. Our results in the unitarity limit are in quite good agreement with quantum Monte Carlo simulation\cite{Gio,ADD} as well as a more self-consistent version of NSR\cite{HuiHu}. 
\par
In Fig. \ref{fig2}, the apparent first-order phase transition in the BEC regime is an artifact of the approximate NSR theory we are using. The reason is as follows. In the BEC regime (where $\mu\ll-\varepsilon_{\rm F}$), the single-particle BCS excitations $E_{\bf p}=\sqrt{\xi_{\bf p}^2+\Delta^2}$ have a large energy gap given by $E_g\equiv\sqrt{\mu^2+\Delta^2}\simeq |\mu|$. This energy gap still exists at $T_{\rm c}$, where $\Delta=0$. In this regime, we can set $\tanh(\beta E_{\rm p}/2)\simeq 1$ in Eq. (\ref{eq.10}) and $f(E_{\rm p})\simeq 0$ in Eqs. (\ref{eq.16})-(\ref{eq.18}). Then, Eq. (\ref{eq.26}) reduces to the expression $\mu=-1/2ma_s^2$, and Eq. (\ref{eq.27}) becomes, by expanding the correlation functions $\Xi_{ij}({\bf q},i\nu_n)$ in powers of ${\bf q}$ and $i\nu_n$ [For the details, see a similar calculation given in Ref. \cite{Ed}.], 
\begin{eqnarray}
{N \over 2}
&=&
N_{c0}
-{1 \over \beta}\sum_{{\bf q},\nu_n}D({\bf q},i\nu_n)e^{i\delta\nu_n}
\nonumber
\\
&=&
N_{c0}
+{1 \over 2}\sum_{\bf q}
\Bigl[
{\varepsilon_{\bf q}^B+U_MN_{c0} \over \omega_{\bf q}}\coth{\beta \over 2}\omega_{\bf q}-1
\Bigr]
\nonumber
\\
&\equiv& 
N_{c0}+N_d.
\label{eq.36}
\end{eqnarray}
Here, $\varepsilon_{\bf q}^B=q^2/2M$ ($M=2m$) and
\begin{equation}
\omega_{\bf q}=\sqrt{\varepsilon^B_{\bf q}(\varepsilon^B_{\bf q}+2U_MN_{c0})}
\label{eq.37}
\end{equation}
is the Bogoliubov excitation spectrum in an interacting gas of Bose molecules. $D$ in Eq. (\ref{eq.36}) is the Bose Green's function, describing Bogoliubov excitations
\begin{equation}
D({\bf q},i\nu_n)=
{
i\nu_n+\varepsilon_{\bf q}^B+U_MN_{c0} 
\over 
(i\nu_n)^2-\omega_{\bf q}^2
},
\label{eq.39}
\end{equation}
where $U_M=4\pi a_M/M$ (where in our theory $a_M=2a_s$, and $M=2m$) is the effective $s$-wave repulsive interaction between Cooper pairs. $N_{c0}$ is given by
\begin{equation}
N_{c0}\equiv
\sum_{\bf p}{\Delta^2 \over 4\xi_{\bf p}^2}
={\sqrt{2}m^{3/2}\Delta^2 \over 16\pi\sqrt{|\mu|}}.
\label{eq.38}
\end{equation}
In Sec. IV, we prove that $N_{c0}$ as defined in Eq. (\ref{eq.38}) corresponds precisely with the formal definition for the condensate fraction in a Fermi superfluid in the BEC limit. $N_d$ as defined in Eq. (\ref{eq.36}) is the number of molecules which are not Bose-condensed, again in the BEC limit. 
\par
Equations (\ref{eq.36}) and (\ref{eq.37}) show that the strong-coupling BEC limit corresponds to the Popov approximation for a weakly-interacting molecular Bose gas\cite{Strinati1}. As is well known (see, for example, Ref. \cite{Shi}), the Popov approximation gives a spurious first-order phase transition at $T_{\rm c}$. This is the origin of the bendover or first-order phase transition evident in Fig. \ref{fig2}\cite{Strinati2} and other figures in this paper. It is well known how to overcome this problem, namely one has to include many-body renormalization effects due to the interaction $U_M$\cite{Stoof,Shi}. Including such higher order corrections past the NSR Gaussian fluctuations considered in this paper is also crucial for determining the correct value of effective interaction $U_M$. The molecular scattering length $a_M=2a_s$ which is obtained in Eq. (\ref{eq.36}) is characteristic of the NSR treatment of fluctuations\cite{Randeria}. The correct result $a_M=0.6a_s$\cite{Petrov} requires going past the Gaussian approximation\cite{Pieli,Stoof}. In this paper, in contrast, we only treat pairing fluctuations within NSR. However, within this approximation, we calculate the superfluid density (and condensate fraction $N_c$ in Sec V) in a consistent manner. 
\par
As discussed in Ref. \cite{Shi}, the Popov approximation becomes invalid in the small region close to $T_{\rm c}$ given by
\begin{equation}
\delta t\equiv {T_{\rm c}-T \over T_{\rm c}}\lesssim 
\Bigl({1 \over 6\pi^2}
\Bigr)^{1/3}
(k_{\rm F}a_M)=0.26(k_{\rm F}a_M).
\label{add2}
\end{equation}
Although we plot numerical results in the present paper in the whole temperature region for completeness, we emphasize that the restriction in Eq. (\ref{add2}) also holds in the BEC regime. We note that the region defined in Eq. (\ref{add2}) becomes narrow as one enters deeper into the BEC regime, simply because the molecular scattering length $a_M\propto a_s$ becomes small. Thus, one obtains $\delta t \lesssim 0.26$ at $(k_{\rm F}a_s)^{-1}=2$ (the case shown in Fig. \ref{fig4}, for example) but $\delta t \lesssim 0.1$ at $(k_{\rm F}a_s)^{-1}=5$. As Fig. \ref{fig4} shows, the bendover occurs over an increasingly small region as we go deeper in the BEC region.
\par
Although Eq. (\ref{eq.36}) was obtained in the strong-coupling BEC regime, we note that the condensate fraction $N_{c}$ in Eq. (\ref{eq.38}) is the mean-field approximation $N_{c}$ for a Fermi superfluid. This is consistent with the result found in Sec. V that the mean-field expression for the condensate fraction $N_c$ is a good approximation even in the strong-coupling BEC regime (at least within NSR). The number of molecules $N_d$ in the {\it non-condensate} given in Eq. (\ref{eq.36}), in contrast, is due to the pairing fluctuations. This is discussed in more detail in Sec. V. 
\par
\section{Superfluid density and the single-particle Green's function}
\vskip2mm
In Ref. \cite{Ed}, our discussion of the superfluid density $\rho_s$ started from the thermodynamic potential $\Omega(v_s)$ in the presence of an imposed superfluid velocity (or phase twist) ${\bf v}_s$. This is given by a simple generalization of $\Omega$ given in Eq. (\ref{eq.19}), which is for the case ${\bf v}_s=0$. Here we give an alternative formulation of $\rho_s$ in terms of how the single-particle Green's function is altered in the presence of a supercurrent. Our numerical calculations in Sec. IV are based on this expression, but it can be proven to be equivalent to the one discussed in Ref. \cite{Ed}. The expression for $\rho_s$ discussed in Ref. \cite{Ed} is convenient for understanding the role of collective pairing fluctuations. The result we obtain in this section gives further insight and is convenient for numerical calculations.
\par
When a supercurrent flows in the $z$-direction with the superfluid velocity $v_s=Q_z/2m$, the supercurrent density $J_z$ is given by
\begin{eqnarray}
J_z=\sum_{{\bf p},\sigma}{p_z \over m}
\langle c_{{\bf p}\sigma}^\dagger c_{{\bf p},\sigma}\rangle
=nv_s+{1 \over \beta}
\sum_{{\bf p},\omega_m}
{p_z \over m}
{\rm Tr}
[{\hat g}({\bf p},i\omega_m)].
\label{eq.3.1}
\end{eqnarray}
Here ${\hat g}({\bf p},i\omega_m)$ is the $2\times 2$-matrix single-particle thermal Green's function in the presence of $v_s$. When the second term in Eq. (\ref{eq.3.1}) is expanded to $O(v_s)$, it can be written as  $J_z=\rho_s v_s$, where the superfluid density $\rho_s$ is defined by
\begin{eqnarray}
\rho_s=n+{2 \over \beta}
\sum_{{\bf p},\omega_m}p_z
{\partial \over \partial Q_z}
{\rm Tr}
[{\hat g}({\bf p},i\omega_m)]_{Q_z\to 0}\equiv n-\rho_n.
\label{eq.3.2}
\end{eqnarray}
The second line defines the normal fluid density $\rho_n$.
\par
We first evaluate the second term in Eq. (\ref{eq.3.2}) in the lowest order mean-field approximation. In this case, the supercurrent state is described by the BCS-type Hamiltonian
\begin{equation}
H=\sum_\sigma\int d{\bf r}
\psi_\sigma^\dagger ({\bf r})
\Bigl[
{{\hat {\bf p}}^2 \over 2m}-\mu
\Bigr]
\psi_\sigma({\bf r})
-\int d{\bf r}
[\Delta({\bf r}) 
\psi^\dagger_\uparrow({\bf r})\psi^\dagger_\downarrow ({\bf r})
+h.c.
],
\label{eq.3.3}
\end{equation}
where $\psi_\sigma({\bf r})$ is a fermion field operator, and
\begin{equation}
\Delta({\bf r})=\Delta e^{i{\bf Q}\cdot{\bf r}},~~~~{\bf Q}=(0,0,Q_z)
\label{eq.3.4}
\end{equation}
is the superfluid order parameter for the current carrying state. In momentum space, Eq. (\ref{eq.3.3}) can be written as
\begin{eqnarray}
H
&=&
\sum_{{\bf p},\sigma}\xi_{\bf p}
c_{{\bf p},\sigma}^\dagger c_{{\bf p},\sigma}
-\Delta
\sum_{\bf p}
[
c_{{\bf p}+{\bf Q}/2,\uparrow}
c_{-{\bf p}+{\bf Q}/2,\downarrow}^\dagger+h.c.
]
\nonumber
\\
&=&
\sum_{\bf p}
{\tilde \Psi}_{\bf p}^\dagger
\Big[{\tilde \xi}_{\bf p}
\tau_3+\alpha_{\bf p}-\Delta\tau_1
\Bigr]
{\tilde \Psi}_{\bf p},
\label{eq.3.5}
\end{eqnarray}
where constant terms have been left out [see Eq. (\ref{eq.3})]. The effects of the supercurrent $v_s$ appear in the Doppler shift term, $\alpha_{\bf p}\equiv {\bf Q}\cdot{\bf p}/2m$, and in ${\tilde \xi}_{\bf p}\equiv \varepsilon_{\bf p}-{\tilde \mu}$, with ${\tilde \mu}\equiv \mu-Q^2/8m$. However, we do not have to take this dependence of ${\tilde \xi}_{\bf p}$ on $\mu$ into account in calculating $\rho_n$ in Eq. (\ref{eq.3.2}) because it is of order of $O(v_s^2)$. ${\tilde \Psi}_{\bf p}$ is the two-component Nambu field in the supercurrent state, 
\begin{eqnarray}
{\tilde \Psi}_{\bf p}=
\left(
\begin{array}{c}
c_{{\bf p}+{\bf Q}/2,\uparrow}\\
c_{-{\bf p}+{\bf Q}/2,\downarrow}^\dagger
\end{array}
\right).
\label{eq.3.6}
\end{eqnarray}
The BCS-Gor'kov Hamiltonian in Eq. (\ref{eq.3.5}) is described by the $2\times 2$-matrix single-particle thermal Green's function with $v_s\ne 0$ which has the well-known form [compare with Eq. (\ref{eq.9})]
\begin{eqnarray}
{\hat g}_0({\bf p},i\omega_m)
\equiv
-\int_0^\beta d\tau
e^{i\omega_m\tau}
\langle
{\rm T}_\tau
\{
{\tilde \Psi}_{\bf p}(\tau)
{\tilde \Psi}^\dagger_{\bf p}(0)
\}
\rangle
={1 \over (i\omega_m-\alpha_{\bf p})-{\tilde \xi}_{\bf p}\tau_3+\Delta\tau_1}.
\label{eq.3.7}
\end{eqnarray}
\par
Substituting Eq. (\ref{eq.3.7}) into Eq. (\ref{eq.3.2}), one obtains the mean-field result
\begin{equation}
\rho_n^F=-{2 \over 3m}\sum_{\bf p}p^2
{\partial f(E_{\bf p}) \over \partial E_{\bf p}}.
\label{eq.46}
\end{equation}
This is the expected Landau expression for the normal fluid density due to Fermi quasiparticles with the BCS spectrum $E_{\bf p}$ when $v_s=0$. This expression gives the Fermi contribution $\rho_n^F$ to the total normal fluid density $\rho_n$ throughout the BCS-BEC crossover, and is not limited to the weak-coupling BCS limit. See, Ref. \cite{Ed} for additional discussion.
\par
In addition to $\rho_n^F$ as given by Eq. (\ref{eq.46}), the correction to ${\hat g}_0$ to first order in $v_s$ gives rise to an additional fluctuation contribution $\rho_n^B$ to the normal fluid density. This should be consistent with the number equation in Eq. (\ref{eq.23}), for the $v_s=0$ state,
\begin{eqnarray}
N
&=&
\sum_{\bf p}1
+{1 \over \beta}\sum_{{\bf p},\omega_m}{\rm Tr}
[\tau_3{\hat G}({\bf p},i\omega_m)],
\label{eq.3.8}
\end{eqnarray}
where the renormalized single-particle Green's function ${\hat G}$
\begin{equation}
{\hat G}({\bf p},i\omega_m)={\hat G}_0({\bf p},i\omega_m)+{\hat G}_0({\bf p},i\omega_m){\hat \Sigma}({\bf p},i\omega_m){\hat G}_0({\bf p},i\omega_m),
\label{eq.3.9b}
\end{equation}
involves the correction from the lowest order quasiparticle self-energy due to coupling with collective fluctuations 
\begin{eqnarray}
{\hat \Sigma}({\bf p},i\omega_m)=
{U \over \beta}\sum_{{\bf q},\nu_n}
{1 \over \eta({\bf q},i\nu_n)}
\Bigl[
(1
&+&U\Xi_{11}({\bf q},i\nu_n))\tau_+{\hat G}_0
({\bf p}+{\bf q},i\omega_m+i\omega_n)\tau_- 
\nonumber
\\
&+&
(1+U\Xi_{22}({\bf q},i\nu_n))\tau_-{\hat G}_0
({\bf p}+{\bf q},i\omega_m+i\omega_n)\tau_+ 
\nonumber
\\
&-&2U\Xi_{12}({\bf q},i\nu_n)\tau_+{\hat G}_0
({\bf p}+{\bf q},i\omega_m+i\omega_n)\tau_+
\Bigr].
\label{eq.3.10}
\end{eqnarray}
Here, $\eta({\bf q},i\nu_n)\equiv {\rm det}[1+U{\hat \Xi}({\bf q},i\nu_n)]$, and $\tau_\pm\equiv (\tau_1\pm i\tau_2)/2$. In obtaining this result, we have carried out the $\mu$-derivative on $\Xi_{ij}$ in Eq. (\ref{eq.23}) by using the identity,
\begin{eqnarray}
{\partial{\hat G}_0 \over \partial \mu}=
-{\hat G}_0\tau_3{\hat G}_0.
\label{eq.3.9}
\end{eqnarray}
For example, 
\begin{eqnarray}
{\partial \Xi_{11} \over \partial \mu}
=
&-&
{1 \over \beta}
\sum_{{\bf p},i\omega_m}
\Bigl[
{\rm Tr}[\tau_3{\hat G}_0({\bf p},i\omega_m)\tau_-{\hat G}({\bf p}+{\bf q},i\omega_m+i\nu_n)\tau_+{\hat G}_0({\bf p},i\omega_m)]
\nonumber
\\
&+&
{\rm Tr}[\tau_3{\hat G}_0({\bf p}+{\bf q},i\omega_m+i\nu_n)\tau_+{\hat G}({\bf p},i\omega_m)\tau_-{\hat G}_0({\bf p}+{\bf q},i\omega_m+i\nu_n)]
\Bigr].
\label{eq.3.8z}
\end{eqnarray}
\par
In the presence of a supercurrent, the Green's function analogous to Eq. (\ref{eq.3.9b}) is given by
\begin{equation}
{\hat g}({\bf p},i\omega_m)={\hat g}_0({\bf p},i\omega_m)+{\hat g}_0({\bf p},i\omega_m){\hat \Sigma}({\bf p},i\omega_m){\hat g}_0({\bf p},i\omega_m),
\label{eq.3.11}
\end{equation}
where ${\hat \Sigma}$ is given by Eq. (\ref{eq.3.10}) but with ${\hat g}_0$ in Eq. (\ref{eq.3.7}) replacing ${\hat G}_0$. Substituting Eq. (\ref{eq.3.11}) into Eq. (\ref{eq.3.2}), we obtain
\begin{equation}
\rho_n=\rho_n^F+\rho_n^B.
\label{eq.3.12}
\end{equation}
\par
The Fermi contribution is given by Eq. (\ref{eq.46}). The bosonic fluctuation contribution $\rho_n^B$ is given by
\begin{eqnarray}
\rho_n^B
&=&{2 \over \beta}\sum_{{\bf p},\omega_m}
p_z {\partial \over \partial Q_z}
{\rm Tr}[{\hat g}_0({\bf p},i\omega_m)
{\hat \Sigma}({\bf p},i\omega_m){\hat g}_0({\bf p},i\omega_m)]_{Q_z\to 0}
\nonumber
\\
&=&
-{2 \over \beta}\sum_{{\bf p},\omega_m}p_z{\partial \over \partial Q_z}
{\rm Tr}
\Bigl[
{\hat \Sigma}({\bf p},i\omega_m)
{\partial {\hat g}_0({\bf p},i\omega_m) \over \partial i\omega_m}
\Bigr]_{Q_z\to 0},
\label{eq.3.15}
\end{eqnarray}
where we have used the identity
\begin{equation}
{\partial {\hat g}_0 \over \partial i\omega_m}=-{\hat g}_0{\hat g}_0.
\label{eq.3.14}
\end{equation}
Substituting Eq. (\ref{eq.3.10}) [with ${\hat G}_0\to {\hat g}_0$] into Eq. (\ref{eq.3.15}), and using the identity
\begin{equation}
{\partial {\hat g}_0 \over \partial \alpha_{\bf p}}=-{\partial {\hat g}_0 \over \partial i\omega_m},
\label{eq.3.16}
\end{equation}
we find
\begin{eqnarray}
\rho_n^B
&=&
-{2 \over \beta^2}
\sum_{{\bf p},\omega_m}{\partial \over \partial Q_z}
\sum_{{\bf q},\nu_n}
p^z_-
{U \over \eta({\bf q},i\nu_n)}
\nonumber
\\
&\times&
{\rm Tr}\Big[
[1+U\Xi_{11}({\bf q},i\nu_n)]\tau_+{\hat g}_0({\bf p}_+,i\omega_m+i\nu_n)\tau_-{\partial {\hat g}_0({\bf p}_-,i\omega_m) \over \partial i\omega_m}
\nonumber
\\
&+&
[1+U\Xi_{22}({\bf q},i\nu_n)]\tau_-{\hat g}_0({\bf p}_+,i\omega_m+i\nu_n)\tau_+{\partial {\hat g}_0({\bf p}_-,i\omega_m) \over \partial i\omega_m}
\nonumber
\\
&-&
2U
\Xi_{12}({\bf q},i\nu_n)\tau_+{\hat g}_0({\bf p}_+,i\omega_m+i\nu_n)
\tau_+{\partial {\hat g}_0({\bf p}_-,i\omega_m) \over \partial i\omega_m}
\Bigr]_{Q_z\to 0}
\nonumber
\\
&=&
-
{2m \over \beta}
{\partial \over \partial Q_z}
\sum_{{\bf q},\nu_n}
{U \over \eta({\bf q},i\nu_n)}
\Bigl[
[1+U\Xi_{11}({\bf q},i\nu_n)]
{\partial \Xi_{22}({\bf q},i\nu_n) \over \partial Q_z}
\nonumber
\\
&+&
[1+U\Xi_{22}({\bf q},i\nu_n)]
{\partial \Xi_{11}({\bf q},i\nu_n) \over \partial Q_z}
-2U\Xi_{12}({\bf q},i\nu_n)
{\partial \Xi_{12}({\bf q},i\nu_n) \over \partial Q_z}
\Bigr]_{Q_z \to 0},
\label{eq.3.17}
\end{eqnarray}
where ${\bf p}_\pm={\bf p}\pm{\bf q}/2$. We note that the correlation functions $\Xi_{ij}({\bf q},i\nu_n)$ appearing in Eq. (\ref{eq.3.17}) are defined as in Eq. (\ref{eq.20}) in terms of the single-particle Green's functions ${\hat g}_0$ in the presence of a supercurrent. In Eq. (\ref{eq.3.17}), the $Q_z$-derivative only acts on $Q_z$ in the Doppler shift term $\alpha_{\bf p}={\bf Q}\cdot{\bf p}/2m$ in ${\hat g}_0$ in Eq. (\ref{eq.3.7}). It does {\it not} act on the $Q_z$ in the shifted chemical potential ${\tilde \mu}=\mu-Q^2/8m$. 
\par
To summarize, we have shown that the total normal fluid density $\rho_n$ associated with fermionic and bosonic degrees of freedom is given by the sum of their contributions:
\begin{eqnarray}
\rho_n
&=&
-
{2 \over 3m}\sum_{\bf p}p^2
{\partial f(E_{\bf p}) \over \partial E_{\bf p}}
-
{2m \over \beta}
{\partial \over \partial Q_z}
\sum_{{\bf q},\nu_n}
{U \over \eta({\bf q},i\nu_n)}
\Bigl[
[1+U\Xi_{11}({\bf q},i\nu_n)]
{\partial \Xi_{22}({\bf q},i\nu_n) \over \partial Q_z}
\nonumber
\\
&+&
[1+U\Xi_{22}({\bf q},i\nu_n)]
{\partial \Xi_{11}({\bf q},i\nu_n) \over \partial Q_z}
-2U\Xi_{12}({\bf q},i\nu_n)
{\partial \Xi_{12}({\bf q},i\nu_n) \over \partial Q_z}
\Bigr]_{Q_z \to 0}.
\label{eq.3.18}
\end{eqnarray}
\par
We note that the superfluid density $\rho_s$ can be also obtained from a current correlation function\cite{Nozieres2,Baym}. The present derivation based on calculating the change in the single-particle Green's function from Eqs. (\ref{eq.3.2}) and (\ref{eq.3.11}) is equivalent to calculating the current correlation function to first order in $v_s$, taking into account both self-energy and current vertex corrections. In Eq. (\ref{eq.3.2}), the $Q_z$-derivative of the first term in Eq. (\ref{eq.3.11}) gives the bare current response function. The $Q_z$-derivative of ${\hat g}_0$ in the second term in Eq. (\ref{eq.3.11}) gives the self-energy corrections, while the $Q_z$-derivative of ${\hat \Sigma}$ gives the vertex corrections. See also Appendix A of Ref. \cite{Ed} for further discussion.
\par
\vskip3mm
\section{Numerical Results for Superfluid Density}
\par
In this section, we present numerical results for the superfluid density $\rho_s$, starting from the expression given in Eq. (\ref{eq.3.18}) in Sec. III. As we have noted earlier, the expression for $\rho_s$ derived in Ref. \cite{Ed} would give identical results. We emphasize that these numerical results for $\rho_s$ use the renormalized values of both $\Delta$ and $\mu$\cite{Ohashi2} which determine the BCS quasiparticle spectrum over the entire BCS-BEC crossover.  
\par
\begin{figure}
\includegraphics[width=8cm,height=10cm]{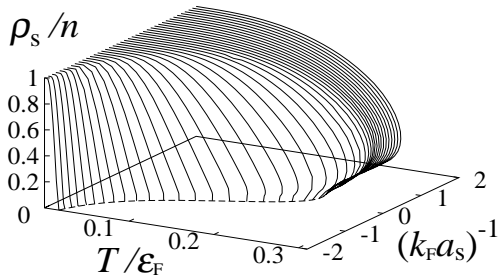}% 
\caption{
Calculated superfluid density $\rho_s$ in the BCS-BEC crossover. The self-consistent solutions for $\Delta$ and $\mu$ shown in Fig. \ref{fig2} are used. The dashed line shows $T_{\rm c}$ [see Fig. \ref{fig3}(a)].
\label{fig6}
}
\end{figure}

Figure \ref{fig6} shows the calculated superfluid density $\rho_s$ in the BCS-BEC crossover. The spurious first-order behavior near $T_{\rm c}$ in the strong-coupling regime is also seen in the self-consistent solutions for $\Delta$ and $\mu$ in Fig. \ref{fig2}. As discussed in Sec. II, this behavior near $T_{\rm c}$ is removed through a more sophisticated treatment of fluctuations, which could lead to the correct second order phase transition. We plot our calculated results for $\rho_s$ close to $T_{\rm c}$, in spite of this problem. In particular, we note that the predicted value of $T_{\rm c}$ in the NSR theory of the BCS-BEC crossover is a good approximation\cite{HuiHu}. The NSR bendover is much less in evidence in the case of a narrow Feshbach resonance, as considered in Ref. \cite{Ohashi2}.
In Appendix C, we prove analytically that our expression in Eq. (\ref{eq.3.18}) gives $\rho_s=n$ at $T=0$ and also that $\rho_s$ vanishes as $\Delta\to 0$ (normal phase). Getting these two limits correctly (as shown in Fig. \ref{fig6}) is very important in any theory of the superfluid density. 
\par

%%%%%%%%%%%%%%%%%%%%%%%%%%%%%%%%%%%%

%%%%%%%%%%%%%%%%%%%%%%%%%%%%%%%%%%%%

%%%%%%%%%%%%%%%%%%%%%%%%%%%%%%%%%%%%

\begin{figure}
\includegraphics[width=8cm,height=6cm]{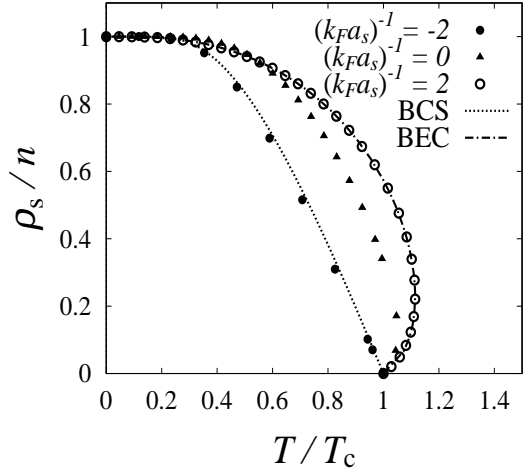}% 
\caption{
Superfluid density $\rho_s$ as a function of temperature in the BCS region (solid circles), unitarity limit (solid triangles) and BEC regime (open circles). `BCS' labels the mean-field BCS result, given by $\rho_s=n-\rho_n^F$ with $\mu=\varepsilon_{\rm F}$. `BEC' gives $\rho_s$ for a dilute Bose gas with $N/2$ bosons described by the excitation spectrum in Eq. (\ref{eq.37}).
\label{fig7}
}
\end{figure}

%%%%%%%%%%%%%%%%%%%%%%%%%%%%%%%%%%%%

Figure \ref{fig7} shows $\rho_s$ as a function of temperature in the BCS regime, unitarity limit, and the BEC regime. We note that $\rho_s$ in the BEC regime [$(k_{\rm F}a_s)^{-1}=2$] is in good agreement with the superfluid density $\rho_s$ of a weakly interacting gas of $N/2$ Bose molecules described by Bogoliubov-Popov excitations in Eq. (\ref{eq.37}), as one expects in the extreme BEC limit. More precisely, in this limit one can show (see Ref. \cite{Ed} for details) that
\begin{equation}
\rho_s=n-\rho_n\simeq n-\rho_n^B,
\label{eq.bec1}
\end{equation}
where $\rho_n^B$ is given by the Landau formula for the normal fluid of an interacting Bose gas
\begin{eqnarray}
\rho_n^B
&=&
-{2 \over 3M}\sum_{\bf q}q^2
{\partial n_B(\omega_{\bf q}) \over \partial \omega_{\bf q}}.
\label{eq.bec2}
\end{eqnarray}
Here, $M=2m$ is the Cooper-pair mass and $n_B(\omega_{\bf q})$ is the Bose distribution function. The excitation energy $\omega_{\bf q}$ is given by Eq. (\ref{eq.37}), calculated with the correct values of $\Delta$ and $\mu$. The curve labeled by BEC in Fig. \ref{fig7} corresponds to the result obtained using Eqs. (\ref{eq.bec1}) and (\ref{eq.bec2}). This shows the importance of a consistent treatment of fluctuation effects in calculating $\rho_s$, $\Delta$, and $\mu$. 
\par
As one approaches the weak-coupling BCS regime, pair fluctuations become weak, so that in this limit the fermionic contribution $\rho_n^F$ becomes dominant. Thus one has
\begin{equation}
\rho_s=n-\rho_n\simeq\rho_n^F,
\label{eq.bec3}
\end{equation}
where $\rho_n^F$ is given by the Landau formula for the normal fluid in Eq. (\ref{eq.46}) with the BCS quasiparticle energies $E_{\bf p}$. The curve labeled by BCS in Fig. \ref{fig7} corresponds to the result obtained using Eqs. (\ref{eq.bec3}) and (\ref{eq.46}).
\par
\vskip3mm
\section{Condensate fraction in the BCS-BEC crossover}
\par
In this section, we calculate the condensate fraction $N_c$ in the BCS-BEC crossover. The condensate fraction $N_c$ in the superfluid phase is most conveniently defined\cite{Yang} as the maximum eigenvalue of the two-particle density matrix, ${\tilde \rho}_2({\bf r},{\bf r}',{\bf r}'',{\bf r}''')\equiv\langle\psi_\uparrow^\dagger({\bf r})\psi^\dagger_\downarrow({\bf r}')\psi_\downarrow({\bf r}'')\psi_\uparrow({\bf r}''')\rangle$, where $\psi_\sigma({\bf r})$ is a fermion field operator. The condensate fraction $N_c$ is given as the maximum eigenvalue, of order $N$. When only one eigenvalue is $O(N)$, we find
\begin{equation}
{\tilde \rho}_2({\bf r},{\bf r}',{\bf r}'',{\bf r}''')=N_c\phi_0^*({\bf r},{\bf r}')\phi_0({\bf r}'',{\bf r}'''),
\label{eq.40}
\end{equation}
where terms of order $O(1)$ have been ignored. Here $\phi_0({\bf r},{\bf r}')$ is the (normalized) two-particle eigenfunction of ${\tilde \rho}_2$ with the eigenvalue $N_c$. The off-diagonal long range order of a Fermi superfluid\cite{Yang} is characterized as, for large separation of (${\bf r},{\bf r}'$) and (${\bf r}'',{\bf r}'''$),
\begin{equation}
{\tilde \rho}_2({\bf r},{\bf r}',{\bf r}'',{\bf r}''')=
\langle\psi_\uparrow^\dagger({\bf r})\psi^\dagger_\downarrow({\bf r}')\rangle
\langle\psi_\downarrow({\bf r}'')\psi_\uparrow({\bf r}''')\rangle.
\label{eq.41}
\end{equation}
Comparing Eqs. (\ref{eq.40}) and (\ref{eq.41}), the condensate fraction $N_c$ is seen to be the normalization factor of the Cooper-pair wavefunction, $\Phi({\bf r},{\bf r}')\equiv \langle\psi_\downarrow({\bf r})\psi_\uparrow({\bf r}')\rangle$ (see, for example, Ref. \cite{Gio}),
\begin{equation}
N_c=\int d{\bf r}d{\bf r}'
|\Phi({\bf r},{\bf r}')|^2.
\label{eq.42}
\end{equation}
In physical terms, the maximum eigenvalue $N_c$ describes the occupancy of two-particle states.
\par
In a uniform Fermi superfluid, the BCS mean-field approximation gives 
\begin{eqnarray}
\Phi({\bf r},{\bf r}')=\sum_{\bf p}
\langle
c^\dagger_{{\bf p}\uparrow}c^\dagger_{-{\bf p}\downarrow}
\rangle
e^{i{\bf p}\cdot({\bf r}-{\bf r}')}
=\sum_{\bf p}
{1 \over 2E_{\bf p}}\tanh{\beta \over 2}E_{\bf p}
e^{i{\bf p}\cdot({\bf r}-{\bf r}')}.
\label{eq.43}
\end{eqnarray}
In the strong-coupling BEC regime (where $\mu\ll-\varepsilon_{\rm F}$), we can set $\tanh{\beta E_{\bf p}/2}=1$ in Eq. (\ref{eq.43}). In this case, substituting Eq. (\ref{eq.43}) into Eq. (\ref{eq.42}), the mean-field expression for the condensate fraction ($\equiv N_{c0}$) in the BEC regime reduces to
\begin{equation}
N_{c0}=\sum_{\bf p}{\Delta^2 \over 4E_{\bf p}^2}\simeq 
\sum_{\bf p}{\Delta^2 \over 4\xi_{\bf p}^2}.
\label{eq.44}
\end{equation}
In obtaining this expression, we have used the fact that $|\mu|\gg\Delta$ in the BEC regime\cite{Engelbrecht}. 
\par
More generally, in terms of the single-particle Green's functions, one can write Eq. (\ref{eq.42}) as
\begin{eqnarray}
N_c
=
{1 \over \beta^2}\sum_{{\bf p},\omega_m,\omega_m'}
G_{21}({\bf p},i\omega_m)
G_{12}({\bf p},i\omega_m').
\label{eq.4.1}
\end{eqnarray}
To calculate the strong-coupling effects on $N_c$, we substitute Eq. (\ref{eq.3.9b}) into Eq. (\ref{eq.4.1}). Since this Green's function ${\hat G}={\hat G}_0+{\hat G}_0{\hat \Sigma}{\hat G}_0$ only includes first order self-energy corrections, we only retain the correction terms to $N_c$ to $O({\hat \Sigma})$, giving
\begin{equation}
N_0=N_{c0}+\delta N_c.
\label{eq.4.2}
\end{equation}
Here, the mean-field component $N_{c0}$ is the BCS Fermi quasiparticle contribution
\begin{eqnarray}
N_{c0}
&\equiv&
{1 \over \beta^2}\sum_{{\bf p},\omega_m,\omega_m'}
G_0^{21}({\bf p},i\omega_m)
G_0^{12}({\bf p},i\omega_m')
\nonumber
\\
&=&
\sum_{\bf p}{\Delta^2 \over 4E_{\bf p}^2}
\tanh^2{\beta E_{\bf p} \over 2}.
\label{eq.4.3}
\end{eqnarray}
The first order fluctuation contribution $\delta N_{c}$ is given by
\begin{eqnarray}
\delta N_{c}
&=&
{1 \over \beta^2}\sum_{{\bf p},\omega_m,\omega_m'}
\Bigl[
G_0^{21}({\bf p},i\omega_m)
{\rm Tr}[\tau_-{\hat G}_0({\bf p},i\omega_m'){\hat \Sigma}({\bf p},i\omega_m'){\hat G}_0({\bf p},i\omega_m')]
\nonumber
\\
&+&
{\rm Tr}[\tau_+{\hat G}_0({\bf p},i\omega_m){\hat \Sigma}({\bf p},i\omega_m){\hat G}_0({\bf p},i\omega_m)]
G_0^{12}({\bf p},i\omega_m')
\Bigr].
\label{eq.4.4}
\end{eqnarray}
\par
\begin{figure}
\includegraphics[width=10cm,height=6cm]{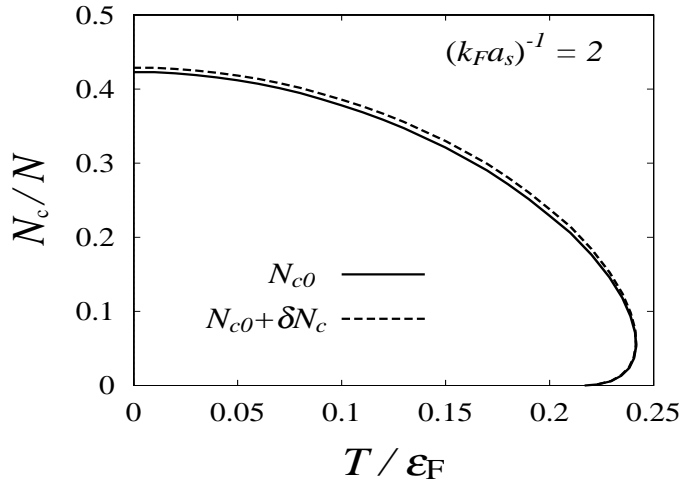}% 
\caption{
Condensate fraction $N_c$ in the strong-coupling BEC regime. The solid line shows $N_{c0}$ and the dashed line includes the small correction $\delta N_c$ from self-energies due to pairing fluctuations.
\label{fig8}
}
\end{figure}

The correction term $\delta N_{c}$ in Eq. (\ref{eq.4.4}) is not important in the weak-coupling BCS regime, where fluctuation effects clearly can be ignored. Figure \ref{fig8} shows that $\delta N_c$ is also negligibly small in the strong-coupling regime. Thus $N_c$ is well approximated by the mean-field expression in Eq. (\ref{eq.4.3}) over the entire BCS-BEC crossover, at least in our NSR-type approximation. We recall that the same pair fluctuations made a large contribution to $\rho_s$ as we went from the BCS to BEC region. The difference is that by definition, $\delta N_c$ in Eq. (\ref{eq.4.4}) arises from self-energy corrections to the single-particle anomalous Green's function $G_{12}$. There is no distinct bosonic contribution, such as $\rho_n^B$ in the normal fluid density. Thus it is not unexpected that the fluctuations are a small correction to $N_c$.
\par
We note, however, since fluctuations in the Cooper channel are taken into account in the equation of state in Eq. (\ref{eq.27}), they modify the condensate fraction $N_c$ given by Eq. (\ref{eq.4.3}). For example, let us consider the BEC regime at $T=0$, where the gap equation gives $\mu=-1/2ma_s^2$. In this limit, the number equation reduces to
\begin{equation}
N=N_{c0}+N_d,
\label{eq.4.5}
\end{equation}
where $N_{c0}$ is given by Eq. (\ref{eq.38}) and [taking the $T\to 0$ limit of Eq. (\ref{eq.36})]
\begin{equation}
N_d=
{1 \over 2}\sum_{\bf q}
\Bigl[
{\varepsilon^B_{\bf q}+U_MN_{c0} \over E_{\bf q}^B}-1
\Bigr]
\simeq {8 \over 3\sqrt{\pi}}(N_{c0}a_M)^{3 \over 2}
\label{eq.4.6}
\end{equation}
gives the quantum depletion from the molecular condensate due to the effective interaction $U_M$ between Cooper pairs. Recently, the mean-field result in Eq. (\ref{eq.4.3}) has been used to study the condensate fraction $N_c$ in a superfluid Fermi gas at $T=0$\cite{Luca}. In this case, the gap and number equation in the mean-field approximation for the BCS-BEC crossover reduce to
\begin{equation}
1=-{4\pi a_s \over m}
\sum_{\bf p}
\Bigl[
{1 \over 2E_{\bf p}}-{1 \over 2\varepsilon_{\bf p}}
\Bigr],
\label{eq.4.7}
\end{equation}
\begin{equation}
N=\sum_{\bf p}
\Bigl[
1-{\xi_{\bf p} \over E_{\bf p}}
\Bigr].
\label{eq.4.8}
\end{equation}
In the BEC regime, while Eq. (\ref{eq.4.7}) again gives $\mu=-1/2ma_s^2$, Eq. (\ref{eq.4.8}) reduces to $N/2=N_{c0}$ in Eq. (\ref{eq.36}). As expected, the depletion $N_d$ at $T=0$ from the condensate due to the interaction between Cooper pairs is omitted when the BCS-BEC crossover is described by mean-field approximation\cite{Luca}.
\par

%%%%%%%%%%%%%%%%%%%%%%%%%%%%%%%%%%%%

%%%%%%%%%%%%%%%%%%%%%%%%%%%%%%%%%%%%

%%%%%%%%%%%%%%%%%%%%%%%%%%%%%%%%%%%%

\begin{figure}
\includegraphics[width=10cm,height=6cm]{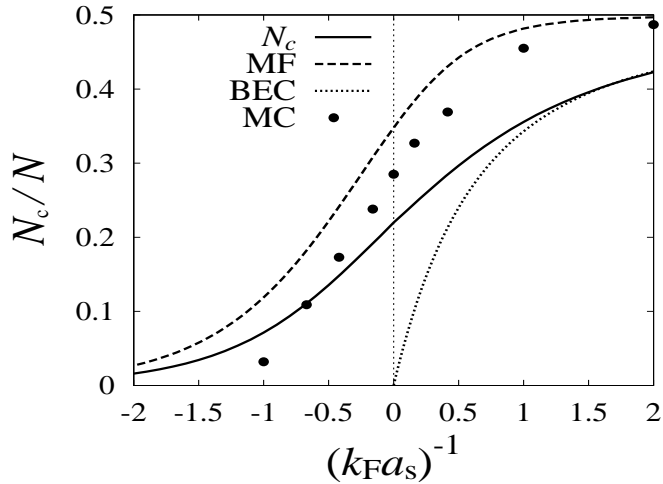}% 
\caption{
Calculated condensate fraction $N_{c}$ in the BCS-BEC crossover at $T=0$ (solid line). In this figure, as well as in Figs. \ref{fig10} and \ref{fig11}, we only show $N_{c0}$. `MF' shows the condensate fraction in the case when $\Delta$ and $\mu$ are determined in the mean-field results in Eqs. (\ref{eq.4.7}) and (\ref{eq.4.8}). `BEC' shows the result for a Bose gas described by Eqs. (\ref{eq.4.5}) and (\ref{eq.4.6}). The solid circles shows recent Monte Carlo results for $N_c$\cite{Gio} for comparison.
\label{fig9}
}
\end{figure}

%%%%%%%%%%%%%%%%%%%%%%%%%%%%%%%%%%%%

Figure \ref{fig9} shows the condensate fraction in the BCS-BEC crossover regime at $T=0$. Because of the omission of the quantum depletion effect, the simple mean-field result (MF) given in Ref. \cite{Luca} is larger than our result ($N_{c0}$). In the region $(k_{\rm F}a_s)^{-1}\gesim 1.5$, our results are well described by the condensate fraction for a superfluid molecular Bose gas given by Eqs. (\ref{eq.4.5}) and (\ref{eq.4.6}) (labeled BEC in Fig. \ref{fig9}). 
\par
Figure \ref{fig9} also compares our $T=0$ results with those obtained by quantum Monte Carlo (MC) simulations\cite{Gio}. The latter calculation gives results consistent with $a_M=0.6a_s$\cite{Petrov}. In contrast, our NSR theory gives the larger mean-field molecular scattering length $a_M=2a_s$. As a result, we overestimate the magnitude of the depletion and thus our values for $N_c$ are smaller than the MC calculation in the BCS-BEC crossover regime. The measurement of the depletion deep in the BEC regime would be a useful way of determining the magnitude of $a_M$.
\par

%%%%%%%%%%%%%%%%%%%%%%%%%%%%%%%%%%%%

\begin{figure}
\includegraphics[width=10cm,height=12cm]{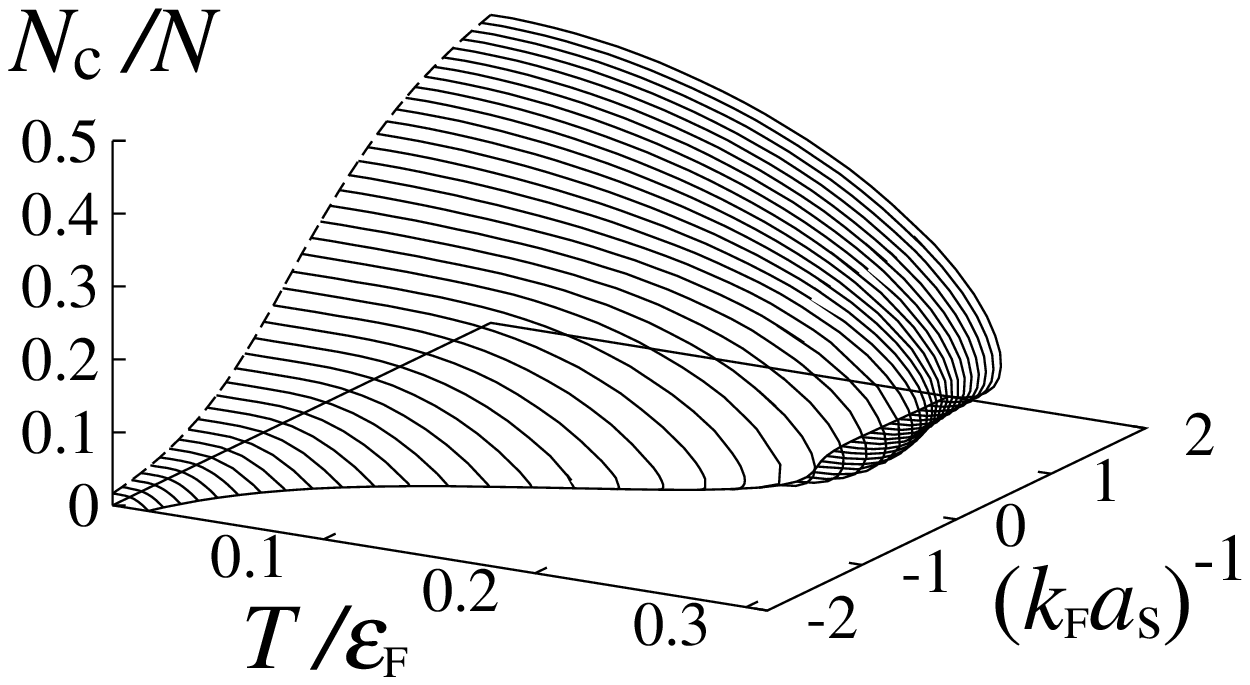}% 
\caption{
Condensate fraction $N_c$ as a function of temperature in the BCS-BEC crossover. In this calculation, the self-consistent solutions for $\Delta$ and $\mu$ in Fig. \ref{fig2} are used.
\label{fig10}
}
\end{figure}

%%%%%%%%%%%%%%%%%%%%%%%%%%%%%%%%%%%%

%%%%%%%%%%%%%%%%%%%%%%%%%%%%%%%%%%%%

\begin{figure}
\includegraphics[width=10cm,height=6cm]{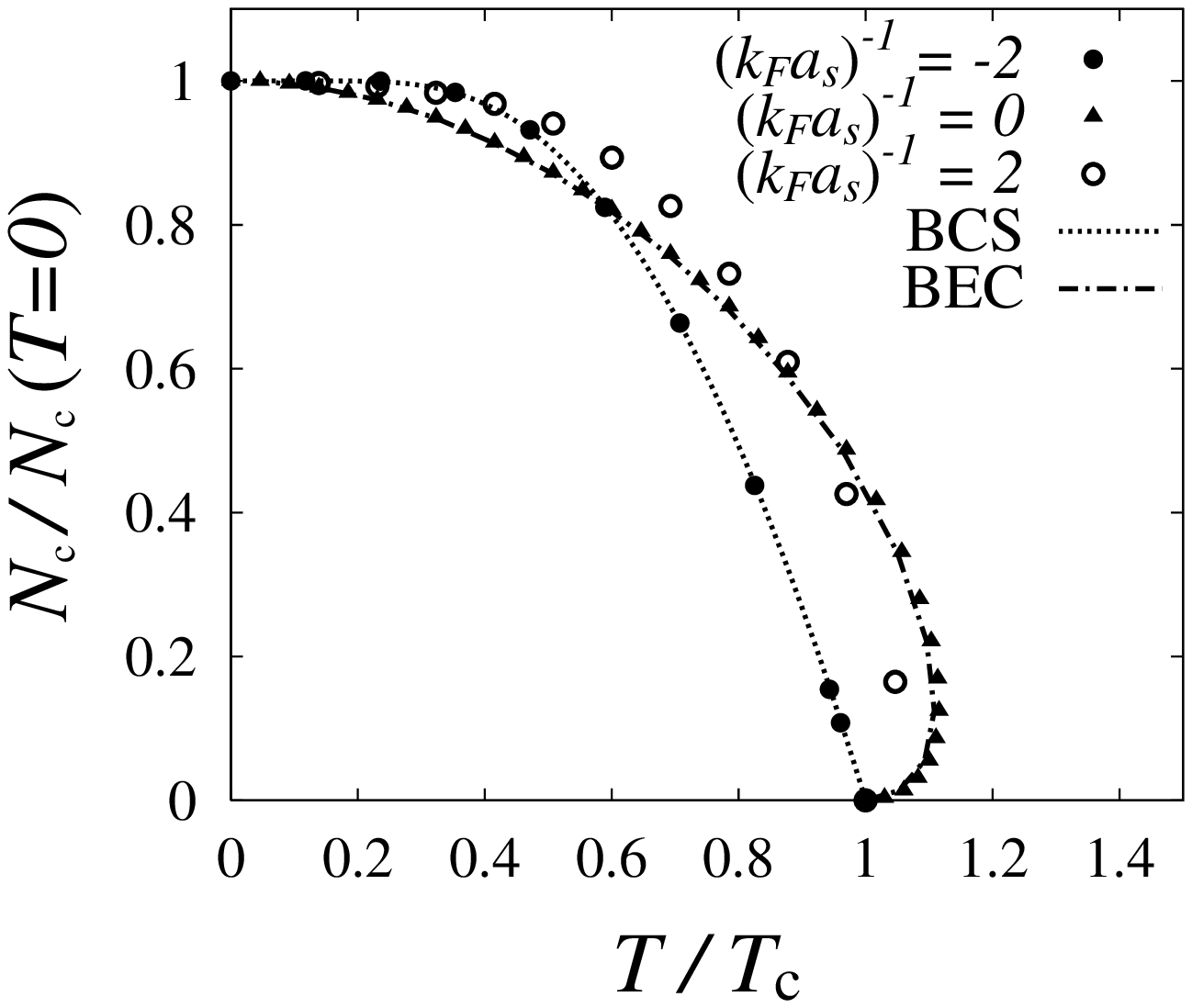}% 
\caption{
Condensate fraction $N_c$ as a function of temperature in the BCS regime, unitarity limit, and the BEC regime.  The curve `BCS' is the weakly-coupling BCS result. The curve labeled `BEC' is the condensate fraction for a Bose superfluid determined by Eq. (\ref{eq.36}). Results are normalized to values at $T=0$.
\label{fig11}
}
\end{figure}

%%%%%%%%%%%%%%%%%%%%%%%%%%%%%%%%%%%%

Figure \ref{fig10} shows the condensate fraction in the BCS-BEC crossover at finite temperatures. In the weak-coupling BCS regime, the condensate fraction $N_c$ is very small even far below $T_{\rm c}$, because only atoms very close to the Fermi surface form Cooper pairs which are Bose-condensed. In this regime, Fig. \ref{fig11} shows that the temperature dependence of $N_c$ is very well described by the weak-coupling BCS result. In the crossover region, the temperature dependence of $N_c$ deviates from the simple BCS result, as shown by the case $(k_{\rm F}a_s)^{-1}=0$ in Fig. \ref{fig11}. In the BEC limit, Fig. \ref{fig10} shows that the condensate fraction at $T=0$ approaches $N/2$, reflecting the fact that all atoms form Cooper pairs which are Bose-condensed. In this regime, Fig. \ref{fig11} shows that the temperature dependence of $N_c$ agrees with the condensate fraction for a dilute Bose gas in the Popov approximation given by Eq. (\ref{eq.36}). Figure \ref{fig9} shows that the BEC picture is a very good approximation when $(k_{\rm F}a_s)^{-1}\gesim 1.5$. The fact that $N_c$ agrees with the Popov theory for a weakly interacting molecular Bose gas shows that the superfluid phase transition in this regime is dominated by the thermal depletion of Cooper pair condensate, and not by the dissociation of Cooper pairs characteristic of the weak-coupling BCS regime.
\par
\section{Conclusions}
\par
In this paper, we have calculated the superfluid density $\rho_s$ and condensate fraction $N_c$ in the BCS-BEC crossover regime of a uniform superfluid Fermi gas at finite temperatures. We have included strong-coupling fluctuation effects on both $\rho_s$ and $N_c$ within a Gaussian approximation. The same fluctuation effects were also taken into account in calculating the superfluid order parameter $\Delta$ and Fermi chemical potential $\mu$ in the BCS-BEC crossover, within the NSR theory\cite{Nozieres,Randeria,Ohashi2}. 
\par
The expression we used to calculate $\rho_s$ was derived from the single-particle Green's function in the presence of supercurrent as given by Eq. (\ref{eq.3.1}), which brings in self-energy corrections due to dynamic pair fluctuations. In this paper, we have concentrated on the explicit numerical calculation of $\rho_s$ within a Gaussian approximation. In contrast, our companion paper\cite{Ed} uses a different (but equivalent) formulation which exhibits the structure of $\rho_s$ in a more direct fashion, in particular, the relation to collective modes. 
\par
Our result for the normal fluid density in Eq. (\ref{eq.3.18}) naturally separates into a mean-field part associated with fermions $\rho_n^F$ and a bosonic pairing fluctuation contribution $\rho_n^B$. As discussed in Ref. \cite{Ed}, $\rho_n^F$ is given by the Landau excitation formula in Eq. (\ref{eq.46}) in the whole BCS-BEC crossover. In the strong-coupling BEC regime, $\rho_n^F$ is negligible and $\rho_n^B$ reduces to the Landau formula for the normal fluid of a Bose gas of tightly bound Copper pairs\cite{Ed}. However, in the region near unitarity, $\rho_n^B$ is not expected to be given by a Landau-type formula because the bosonic pairing fluctuations are strongly damped. It is in this region that the numerical calculations for $\rho_s$ reported in this paper are especially useful. 
\par
The superfluid density is a fundamental quantity in two-fluid hydrodynamics\cite{Ed2} and we will use our results in a future study of hydrodynamic modes in the BCS-BEC crossover regime of a Fermi superfluid at finite temperatures. 
\par
In contrast to the superfluid density, the mean-field expression for the condensate fraction $N_c$ is a good approximation even in the strong-coupling BEC regime. The fluctuation contribution to $N_c$ gives rise to the non-condensate component. In the BEC regime, we showed that the fluctuation contribution gives the condensate depletion $N_d$ due to the effective interaction $U_M$ between Cooper pairs, which is finite even at $T=0$.
\par
In the BEC regime, the strong coupling theory presented in this paper reduces to that of a weakly interacting Bose gas of molecules, with an excitation spectrum given by the Bogoliubov-Popov approximation. This is also the origin of the spurious first-order phase transition our theory exhibits (see Figs. \ref{fig6} and \ref{fig7}). This is a well-known problem in dealing with dilute Bose gases\cite{Shi}. The recovery of the second order phase transition in the entire BCS-BEC crossover, as well as the normalized magnitude of the effective interaction between Cooper pairs\cite{Petrov}, both require the inclusion of higher order fluctuations past the NSR Gaussian approximation which we have used. In this regard, we have emphasized that calculating the value of $\rho_s$ is very dependent on using the strong-coupling approximation for $\Delta$ and $\mu$ as well, quantities which determine the single-particle excitation spectrum. Thus, when we calculate $\Delta$ and $\mu$ beyond the Gaussian fluctuation level, we also need to improve the microscopic model used to calculate $\rho_s$. The approach presented in this paper, as well as in Ref. \cite{Ed}, can be the starting point for such improved calculations. 

\vskip2mm
\acknowledgments
N. F. and Y. O. would like to thank H. Matsumoto for useful discussions. Y. O. was financially supported by a Grant-in-Aid for Scientific research from the Ministry of Education, Culture, Sports, Science and Technology of Japan (16740187, 17540368, and 18043005). The work of E. T. and A. G. were supported by NSERC of Canada.
\par 
%
%
%%%%%%%%%%%%%%%%%%%%%%%%%%%%%%%%%%%%
\newpage
%%%%%%%%%%%%%%%%%%%%%%%%%%%%%%%%%%%%

\appendix
\section{Number equation in the BCS-BEC crossover}
We briefly review some of the formalism developed in Ref. \cite{Ohashi2}. The thermodynamic potential $\Omega=\Omega^F+\Omega^B$ consists of the mean-field part $\Omega^F$ due to Fermi quasiparticles and the correction term $\Omega^B$ due to boson fluctuations. The mean-field contribution $\Omega^F$ is given by\cite{Ohashi2,HuiHu}
\begin{eqnarray}
\Omega^F={\Delta^2 \over U}+\sum_{\bf p}[\xi_{p}-E_{\bf p}]
-{2 \over \beta}\sum_{\bf p}\ln [1+e^{-\beta E_{\bf p}}].
\label{eq.11}
\end{eqnarray}
In the NSR Gaussian approximation, the fluctuation contribution is the sum of diagrams $\Omega^B$ shown in Fig. \ref{fig1}, namely

\begin{eqnarray}
\Omega^B=
{1 \over 2\beta}\sum_{{\bf q},\nu_n}
\ln \det 
\left[
1+U\hat{\Xi}({\bf q},i\nu_n)
\right],
\label{eq.19}
\end{eqnarray}
where
\begin{eqnarray}
{\hat \Xi}({\bf q},i\nu_n)
=
{1 \over 4}
\left(
\begin{array}{cc}
\Pi_{11}^0+\Pi_{22}^0+i(\Pi_{12}^0-\Pi_{21}^0) &
\Pi_{11}^0-\Pi_{22}^0 \\
\Pi_{11}^0-\Pi_{22}^0 &
\Pi_{11}^0+\Pi_{22}^0-i(\Pi_{12}^0-\Pi_{21}^0)
\end{array}
\right).
\label{eq.20}
\end{eqnarray}
The correlation functions $\Pi_{ij}^0$ are given within the mean-field approximation by
\begin{eqnarray}
\Pi^0_{ij}({\bf q},i\nu_n)
&=&-\int_{0}^{\beta} d\tau e^{i\nu_n\tau}
\langle
T_{\tau}
\left\{
\rho_{i,{\bf q}}(\tau)\rho_{j,-{\bf q}}(0)
\right\}
\rangle
\nonumber
\\
&=&
{1 \over \beta}
\sum_{{\bf p},\omega_{n}}
T_{\tau}
\left[\tau_{i}{\hat G}_0({\bf p+q}/2,i\omega_{n}+i\nu_{n})
\tau_{j}{\hat G}_0({\bf p-q}/2,i\omega_{n})\right].
\label{eq.14}
\end{eqnarray}
$\Pi^0_{11}$ and $\Pi^0_{22}$ describe the amplitude fluctuations and phase fluctuations of the order parameter, respectively, while $\Pi_{12}^0$ and $\Pi_{21}^0$ are the coupling of these fluctuations. Doing the Fermi Matsubara frequency sum over $\omega_m$, one finds\cite{Ohashi2,Takada}
\begin{eqnarray}
\Pi_{11}^0({\bf q},i\nu_{n})
&=&
\sum_{\bf p}
\left(
1-{\xi_{{\bf p+q}/2}\xi_{{\bf p-q}/2}-\Delta^2 \over E_{{\bf p+q}/2}E_{{\bf p-q}/2}}
\right)
{E_{{\bf p+q}/2}-E_{{\bf p-q}/2} \over (E_{{\bf p+q}/2}-E_{{\bf p-q}/2})^2+\nu_n^2}
\nonumber
\\
&\times&
[f(E_{{\bf p+q}/2})-f(E_{{\bf p-q}/2})]
\nonumber
\\
&-&
\sum_{\bf p}
\left(
1+{\xi_{{\bf p+q}/2}\xi_{{\bf p-q}/2}-\Delta^2 \over E_{{\bf p+q}/2}E_{{\bf p-q}/2}}
\right)
{E_{{\bf p+q}/2}+E_{{\bf p-q}/2} \over (E_{{\bf p+q}/2}+E_{{\bf p-q}/2})^2+\nu_n^2}
\nonumber
\\
&\times&
[1-f(E_{{\bf p+q}/2})-f(E_{{\bf p-q}/2})],
\label{eq.16}
\end{eqnarray}

\begin{eqnarray}
\Pi_{22}^0({\bf q},i\nu_{n})
&=&
\sum_{\bf p}
\left(
1-{\xi_{{\bf p+q}/2}\xi_{{\bf p-q}/2}+\Delta^2 \over E_{{\bf p+q}/2}E_{{\bf p-q}/2}}
\right)
{E_{{\bf p+q}/2}-E_{{\bf p-q}/2} \over (E_{{\bf p+q}/2}-E_{{\bf p-q}/2})^2+\nu_n^2}
\nonumber
\\
&\times&
[f(E_{{\bf p+q}/2})-f(E_{{\bf p-q}/2})]
\nonumber
\\
&-&
\sum_{\bf p}
\left(
1+{\xi_{{\bf p+q}/2}\xi_{{\bf p-q}/2}+\Delta^2 \over E_{{\bf p+q}/2}E_{{\bf p-q}/2}}
\right)
{E_{{\bf p+q}/2}+E_{{\bf p-q}/2} \over (E_{{\bf p+q}/2}+E_{{\bf p-q}/2})^2+\nu_n^2}
\nonumber
\\
&\times&
[1-f(E_{{\bf p+q}/2})-f(E_{{\bf p-q}/2})],
\label{eq.17}
\end{eqnarray}

\begin{eqnarray}
\Pi_{12}^0({\bf q},i\nu_{n})
&=&
\sum_{\bf p}
\left(
{\xi_{{\bf p+q}/2} \over E_{{\bf p+q}/2}}-{\xi_{{\bf p-q}/2} \over E_{{\bf p-q}/2}}
\right)
{\nu_{n} \over (E_{{\bf p+q}/2}-E_{{\bf p-q}/2})^2+\nu_{n}^2}
\nonumber
\\
&\times&
[f(E_{{\bf p+q}/2})-f(E_{{\bf p-q}/2})]
\nonumber
\\
&-&
\sum_{{\bf p}}
\left(
{\xi_{{\bf p+q}/2} \over E_{{\bf p+q}/2}}+{\xi_{{\bf p-q}/2} \over E_{{\bf p-q}/2}}
\right)
{\nu_{n} \over (E_{{\bf p+q}/2}+E_{{\bf p-q}/2})^2+\nu_{n}^2}
\nonumber
\\
&\times&
[1-f(E_{{\bf p+q}/2})-f(E_{{\bf p-q}/2})]
\nonumber
\\
&=&
-\Pi_{21}^0({\bf q},i\nu_n).
\label{eq.18}
\end{eqnarray}
We note that for zero superfluid flow ($v_s=0$), the matrix elements $M_{ij}$ of the inverse fluctuation propagator discussed in Ref. \cite{Ed} are related to $\Xi_{ij}$ by
\begin{equation}
M_{ij}=1+U\Xi_{ij}.
\label{app100}
\end{equation}
In contrast to Ref. \cite{Ed}, Eq. (\ref{eq.20}) splits the fluctuations explicitly into phase and amplitude components.
\par
In calculating $N=-\partial \Omega /\partial\mu$,  we note that, in principle, both $\xi_{\bf p}=\varepsilon_{\bf p}-\mu$ and $\Delta$ depend on $\mu$. Thus, one needs to calculate $\partial \Omega/\partial\Delta=\partial \Omega^F/\partial\Delta+\partial \Omega^B /\partial \Delta$. The first term $\partial \Omega^F/\partial\Delta$ vanishes by definition, since the gap equation (\ref{eq.10}) is the saddle point solution\cite{Randeria}. As noted in our previous paper\cite{Ohashi2}, strictly speaking, the second term $\partial \Omega^B /\partial\Delta$ is a higher order correction within the NSR Gaussian treatment of fluctuations. Thus in a consistent theory built on treating fluctuations to quadratic order, we should omit the term $\partial\Omega^B/\partial\Delta$. Within this approximation, we obtain the expression for $N$ given in Eq. (\ref{eq.23}). If we retained terms involving $\partial \Omega^B /\partial\Delta$ in Eq. (\ref{eq.6}), the resulting equation for $N$ would not reduce to the correct result at $T_{\rm c}$ in the limit $\Delta\to 0$. On the other hand, the importance of this class of higher order correction terms has been pointed out in Refs. \cite{Keeling,HuiHu}. In particular, Ref. \cite{HuiHu} has shown that the correct molecular scattering length in the BEC regime ($a_M=0.6a_s$) is obtained when this term is included. See Ref. \cite{Ed} for additional discussion.
\par

\section{Relation to the coupled fermion-boson model}
\par
In this appendix, we compare the results for a single-channel model given by Eqs. (\ref{eq.26}) and (\ref{eq.27}) with our previous results for the coupled fermion-boson (CFB) model\cite{Ohashi2}. The CFB two-channel model involves treating the Feshbach resonance explicitly by the Hamiltonian,
\begin{eqnarray}
H_{\rm CFB}
&=&\sum_{{\bf p},\sigma}[\varepsilon_{\bf p}-\mu]
c_{{\bf p}\sigma}^\dagger c_{{\bf p}\sigma}
-U_{\rm bg}\sum_{{\bf p},{\bf p}',{\bf q}}
c_{{\bf p}+{\bf q}\uparrow}^\dagger c_{{\bf p}'-{\bf q}\downarrow}^\dagger
c_{{\bf p}'\downarrow} c_{{\bf p}\uparrow}
\nonumber
\\
&+&\sum_{\bf q}[\varepsilon_{\bf q}^B+2\nu-2\mu]b_{\bf q}^\dagger b_{\bf q}
+g_{\rm r}
\sum_{{\bf p},{\bf q}}
[
b_{\bf q}^\dagger 
c_{{\bf p}+{\bf q}/2,\downarrow} c_{-{\bf p}+{\bf q}/2,\uparrow}
+h.c.
].
\label{eq.28}
\end{eqnarray}
Here, $b_{\bf q}^\dagger$ is the creation operator of a molecular boson associated with the Feshbach resonance, with the kinetic energy $\varepsilon^B_{\bf q}\equiv q^2/4M$. The threshold energy $2\nu$ of the Feshbach resonance can be tuned by adjusting a small external magnetic field. $g_{\rm r}$ describes the Feshbach coupling between atoms and a molecule. $U_{bg}$ is a nonresonant weak-interaction, which is taken to be attractive in Eq. (\ref{eq.28}). Since one molecule is a bound state of two Fermi atoms (in different hyperfine states), we take $M=2m$ and must impose a conservation law for the total number of Fermi atoms. The latter constraint has been taken into account in Eq. (\ref{eq.28}) by taking the chemical potential for the molecules to be $2\mu$. 
\par
In a previous paper\cite{Ohashi2}, we used this two-channel CFB model to treat the superfluid properties in the BCS-BEC crossover region within the NSR theory. The resulting coupled equations corresponding to Eqs. (\ref{eq.10}) and (\ref{eq.23}) are given by
\begin{equation}
1=U_{\rm eff}\sum_{\bf p}{1 \over 2E_{\bf p}}\tanh{\beta \over 2}E_{\bf p},
\label{eq.29}
\end{equation}
\begin{eqnarray}
N=2\phi_m^2+N_F^0+2N_B^0-{1 \over 2\beta}
{\partial \over \partial\mu}
\sum_{{\bf q},\nu_n}
\ln
{\rm det}
\Bigl[
1+[U-g_{\rm r}^2{\hat D}^0({\bf q},i\nu_n)]{\hat \Xi}({\bf q},i\nu_n)
\Bigr].
\label{eq.30}
\end{eqnarray}
Here, $U_{\rm eff}\equiv U_{\rm bg}+g_{\rm r}^2/(2\nu-2\mu)$ is an effective pairing interaction associated with the Feshbach resonance. $\phi_m\equiv \langle b_{{\bf q}=0}\rangle$ is the BEC order parameter and $\phi_m^2$ is the number of Bose-condensed Feshbach molecules. Because of the resonance coupling between atoms and molecules, $\phi_m$ is related to the BCS order parameter $\Delta$ by the relation $\phi_m=-g_{\rm r}\Delta/U(2\nu-2\mu)$\cite{Ohashi2}. $N_B^0=\sum_{\bf q}n_B(\varepsilon_{\bf q}^B+2\nu-2\mu)$ gives the thermal occupation of the non-condensed molecules, where $n_B(x)$ is the Bose distribution function. ${\hat D}^0({\bf q},i\nu_n)^{-1}\equiv i\nu_n\tau_3-(\xi_{\bf q}^B+2\nu-2\mu)$ is the $2\times 2$-matrix single-particle Green's function for a free Bose gas of molecules. In Eq. (\ref{eq.29}), the BCS single-particle excitation spectrum $E_{\rm p}$ is given by $E_{\bf p}=\sqrt{\xi_{\bf p}^2+{\tilde \Delta}^2}$, where $\xi_{\bf p}\equiv\varepsilon_{\bf p}-\mu$ and ${\tilde \Delta}\equiv\Delta-g_{\rm r}\phi_m$ is the composite order parameter involving the $\Delta$ and $\phi_m$ coupled order parameters. For more details, see Ref. \cite{Ohashi2}.
\par
Let use now consider the case of a broad Feshbach resonance, where the coupling $g_{\rm r}$ is very large ($g_{\rm r}\sqrt{n}\gg\varepsilon_{\rm F}$). In this limit, the effective pairing interaction $U_{\rm eff}$ defined below Eq. (\ref{eq.30}) can be strong even when the threshold energy $2\nu$ is still much larger than the chemical potential $\mu$ ($\lesssim \varepsilon_{\rm F}$). In this case, one can neglect $\mu$ in $U_{\rm eff}$ in the interesting BCS-BEC crossover regime. Since $2\nu$ is the lowest excitation energy of Feshbach molecules, we can also neglect $2\phi_m^2$ and $2N_B^0$ in Eq. (\ref{eq.30}) when $2\nu\gg2\varepsilon_{\rm F}$. Similarly, for $2\nu\gg2\varepsilon_{\rm F}$, dynamical effects of Feshbach molecules are not important, and we can use the approximation ${\hat D}^0({\bf q},i\nu_n)\simeq {\hat D}^0(0,0)=-1/(2\nu-2\mu)\simeq -1/2\nu$ in Eq. (\ref{eq.30}). The end result is that the coupled equations (\ref{eq.29}) and (\ref{eq.30}) for the two-channel model reduce to a form analogous to Eqs. (\ref{eq.10}) and (\ref{eq.23}) in the single-channel model. We need only replace $\Delta$ by ${\tilde \Delta}$\cite{note3} and $U$ by the two-particle effective interaction associated with the Feshbach resonance,
\begin{equation}
U_{\rm eff}^{2b}\equiv U_{\rm bg}+{g_{\rm r}^2 \over 2\nu}.
\label{eq.31}
\end{equation}
\par
Thus we see how the single-channel description of the BCS-BEC crossover described by Eqs. (\ref{eq.10}) and (\ref{eq.23}) is a special case of the two-channel model in the case of a broad Feshbach resonance $2\nu\gg2\varepsilon_{\rm F}$. The single-channel scattering length $a_s$ is related to the two-channel CFB model parameters by
\begin{equation}
{4\pi a_s \over m}=-{\tilde U}_{\rm bg}-{{\tilde g}_{\rm r}^2 \over 2{\tilde \nu}}.
\label{eq.32}
\end{equation} 
Here, ${\tilde U}_{\rm bg}$, ${\tilde g}_{\rm r}$, and ${\tilde \nu}$ are all renormalized quantities, given by\cite{Ohashi3}
\begin{equation}
{\tilde U}_{\rm bg}\equiv{U \over 1-U\sum_{\bf p}^{\omega_c}{1 \over 2\varepsilon_{\bf p}}},
~~~
{\tilde g}_{\rm r}\equiv{g_{\rm r} \over 1-U\sum_{\bf p}^{\omega_c}{1 \over 2\varepsilon_{\bf p}}},
~~~
2{\tilde \nu}\equiv2\nu-g_{\rm r}^2{\sum_{\bf p}^{\omega_c}{1 \over 2\varepsilon_{\bf p}} \over 1-U\sum_{\bf p}^{\omega_c}{1 \over 2\varepsilon_{\bf p}}}.
\label{eq.35}
\end{equation}
The renormalized unitarity limit ($a_s\to\pm\infty$) corresponds to $2{\tilde \nu}=0$.
\par
\section{Analytical Results at $T=0$ and $\Delta\to 0$}
\par
We first prove that the normal fluid density as given by the expression in Eq. (\ref{eq.3.18}) vanishes at $T=0$. At $T=0$, clearly the first term $\rho_n^{F}$ vanishes. For the fluctuation part $\rho_n^B$, we find from explicit calculations that one can change the $Q_z$-derivative into the $\nu_n$-derivative. Then, we find
\begin{eqnarray}
\rho_n^B(T=0)
&=&-{2 \over \beta}
\sum_{{\bf q},\nu_n}
{q_z^2 \over 2m}
{\partial^2 \over \partial (i\nu_n)^2}
\ln {\rm det}
\Bigl[
1+U{\hat \Xi}({\bf q},i\nu_n)
\Bigr]
\nonumber
\\
&=&
-{2 \over \pi}\sum_{\bf q}{q_z^2 \over 2m}
\int_{-\infty}^\infty dz
n_B(z) {\rm Im}
\Bigl[
{\partial^2 \over \partial z^2}
\ln {\rm det}
[1+U{\hat \Xi}({\bf q},i\nu_n\to z+i\delta)]
\Bigr]
\nonumber
\\
&=&
{2 \over \pi}
\sum_{\bf q}
{q_z^2 \over 2m}
{\rm Im}
\Bigl[
{\partial \over \partial z}
\ln {\rm det}
[1+U{\hat \Xi}({\bf q},z+i\delta)
\Bigr]_{z=-\infty}^{z=0}.
\label{eq.3.19}
\end{eqnarray}
In obtaining the last line, we have used that $n_B(z)=-\theta(-z)$ at $T=0$. One can easily show that Eq. (\ref{eq.3.19}) vanishes, by noting that $\Xi_{ij}(z\to-\infty)=0$ and $\Pi_{11}(z=0)$, $\Pi_{22}(z=0)$, and $\Pi_{12}(z=0)\Pi_{21}(z=0)$ are all real quantities. Thus, we have shown that in our microscopic model, $\rho_n=\rho_n^F+\rho_n^B=0$ at $T=0$, or $\rho_s=n$.
\par
We next show that $\rho_s$ vanishes in the limit $\Delta\to 0$, i.e., in the normal phase above $T_{\rm c}$. To see this, we note that the mean-field part $\rho_n^{F}$ reduces to the number of Fermi atoms when $\Delta\to 0$, namely,
\begin{equation}
\rho_n^{F}(T_{\rm c})=2\sum_{\rm p}f(\xi_{\bf p}). 
\label{eq.3.20}
\end{equation}
For the fluctuation contribution $\rho_n^B$, since $\Xi_{12}=\Xi_{21}=g_0^{12}=g_0^{21}=0$ in the limit $\Delta\to 0$, Eq. (\ref{eq.3.17}) reduces to
\begin{eqnarray}
\rho_n^B(T_{\rm c})
&=&
-{2 \over \beta^2}
\sum_{{\bf p},\omega_m}
{\partial \over \partial Q_z}
\sum_{{\bf q},\nu_n}
p^z_-
{U \over \eta({\bf q},i\nu_n)}
\nonumber
\\
&\times&
\Bigl[
[1+U\Xi_{11}({\bf q},i\nu_n)]
g_0^{22}({\bf p}_+,i\omega_m+i\nu_n)
{\partial g_0^{11}({\bf p}_-,i\omega_m) \over \partial i\omega_m}
\nonumber
\\
&+&
[1+U\Xi_{22}({\bf q},i\nu_n)]
g_0^{11}({\bf p}_+,i\omega_m+i\nu_n)
{\partial g_0^{22}({\bf p}_-,i\omega_m) \over \partial i\omega_m}
\Bigr]_{Q_z\to 0}.
\label{eq.3.21}
\end{eqnarray}
Carrying out the $Q_z$-derivative, we obtain
\begin{eqnarray}
\rho_n^B(T_{\rm c})
&=&
-{1 \over \beta^2}
\sum_{{\bf p},\omega_m}
\sum_{{\bf q},\nu_n}
{U \over \eta({\bf q},i\nu_n)}
\Bigl[
[1+U\Xi_{11}({\bf q},i\nu_n)]
G_0^{22}({\bf p}_+,i\omega_m+i\nu_n)
{\partial G_0^{11}({\bf p}_-,i\omega_m) \over \partial i\omega_m}
\nonumber
\\
&-&
[1+U\Xi_{22}({\bf q},i\nu_n)]
G_0^{11}({\bf p}_+,i\omega_m+i\nu_n)
{\partial G_0^{22}({\bf p}_-,i\omega_m) \over \partial i\omega_m}
\Bigr]
\nonumber
\\
&=&
{1 \over \beta^2}
\sum_{{\bf p},\omega_m}
\sum_{{\bf q},\nu_n}
{U \over \eta({\bf q},i\nu_n)}
[1+U\Xi_{22}({\bf q},i\nu_n)]
\nonumber
\\
&\times&
\Bigl[
G_0^{11}({\bf p}_+,i\omega_m+i\nu_n)
{\partial G_0^{22}({\bf p}_-,i\omega_m) \over \partial i\omega_m}
-
G_0^{22}({\bf p}_-,i\omega_m)
{\partial G_0^{11}({\bf p}_+,i\omega_m+i\nu_n) \over \partial i\omega_m}
\Bigr].
\nonumber
\\
\label{eq.3.22}
\end{eqnarray}
Noting that $\eta({\bf q},i\nu)=[1-U\Xi_{11}({\bf q},i\nu_n)][1-U\Xi_{22}({\bf q},i\nu_n)]$ and $\Xi_{11}({\bf q},i\nu_n)=-\Pi({\bf q},i\nu_n)$ when $\Delta=0$, and using the identity,
\begin{eqnarray}
{1 \over \beta}\sum_{{\bf p},\omega_m}
\Bigl[
G_0^{11}({\bf p}_+,i\omega_m+i\nu_n)
{\partial G_0^{22}({\bf p}_-,i\omega_m) \over \partial i\omega_m}
&-&
G_0^{22}({\bf p}_-,i\omega_m)
{\partial G_0^{11}({\bf p}_+,i\omega_m+i\nu_n) 
\over \partial i\omega_m}
\Bigr]
\nonumber
\\
&=&
-{\partial \over \partial\mu}
{1 \over \beta}\sum_{{\bf p},\omega_m}[G_0^{11}({\bf p}_+,i\omega_m+i\nu_n) G_0^{22}({\bf p}_-,i\omega_m)]
\nonumber
\\
&=&
{\partial \over \partial \mu}\Pi({\bf q},i\nu_n),
\label{eq.3.23}
\end{eqnarray}
one may reduce Eq. (\ref{eq.3.22}) to
\begin{eqnarray}
\rho_n^B(T_{\rm c})=
-{1 \over \beta}\sum_{{\bf q},\nu_n}
{\partial \over \partial \mu}
\ln [1-U\Pi({\bf q},i\nu_n)].
\label{eq.3.24}
\end{eqnarray}
The sum of Eqs. (\ref{eq.3.20}) and (\ref{eq.3.24}) equals the total number of Fermi atoms $n$ as given in Eq. (\ref{eq.6}). Therefore, we have proven that $\rho_n=n$, or $\rho_s=0$ in the limit $\Delta\to 0$ (normal phase). This is an important requirement of any theory of $\rho_s$.

%%%%%%%%%%%%%%%%%%%
\newpage
%%%%%%%%%%%%%%%%%%%

%%%%%%%%%%%%%%%%%%%%%%%%%%%%%%%%%%%%%%%%%%%%%%%%%%%%%%%%%%%%%%%%%%%%

%

\begin{thebibliography}{99}
\bibitem{notez} In all Bose frequency sums in this paper, one needs the convergence factor $e^{i\delta\nu_n}$. This is left implicit.
\bibitem{note1} Since we take $\Delta$ to be real, we need to eliminate the expectation value from $\rho_{1,{\bf q}}$ as $\rho_{1,{\bf q}}-\langle\rho_{1,{\bf q=0}}\rangle\delta_{{\bf q},0}$, although this is not explicitly written in Eq. (\ref{eq.4}).
\bibitem{note3} In a broad Feshbach resonance, although $\phi_m$ is small, the Feshbach coupling $g_{\rm r}$ is large. As a result, the molecular component $g_{\rm r}\phi_m$ in the composite order parameter ${\tilde \Delta}=\Delta-g_{\rm r}\phi_m$ still has a relatively large contribution even in the crossover regime. For further discussion, see Ref. \cite{Ohashi2}.
\end{thebibliography}
\end{document}